\documentclass[twocolumn, pre, showpacs, english, preprintnumbers, amsmath, amssymb, superscriptaddress, aps,longbibliography]{revtex4-2}
\usepackage[subpreambles=true]{standalone}
\usepackage{comment}
\usepackage[utf8]{inputenc}
\usepackage{graphicx}
\usepackage{grffile}
\usepackage[usenames,dvipsnames]{xcolor}
\usepackage{amsmath}
\usepackage[normalem]{ulem}
\usepackage[resetlabels, labeled]{multibib}
\usepackage{appendix}
\newcites{S}{References Supplementary Materials}
\definecolor{orange}{rgb}{1,0.5,0}
\definecolor{goodgreen}{rgb}{0.1,0.5,0}
\definecolor{goodred}{rgb}{0.7,0,0}
\usepackage{lineno}
\renewcommand\vec{\boldsymbol}
\newcommand{\TK}[1]{{\color{CornflowerBlue}[TK: #1]}}

\usepackage{tikz}
\usepackage{tocvsec2}
\usepackage{scrextend}
\makeatletter

\usepackage[colorlinks,urlcolor=goodgreen,citecolor=blue,linkcolor=goodred]{hyperref}

\usepackage{color}

\usepackage{float}
\graphicspath{ {figures/} }

\let\oldepsilon\epsilon \let\epsilon\varepsilon \let\varepsilon\oldepsilon

\makeatother

\usepackage{babel}

\begin{document}

\title{Proximity effect of time-reversal symmetry broken non-centrosymmetric superconductors.}
\newcommand{\orcid}[1]{\href{https://orcid.org/#1}{\includegraphics[width=8pt]{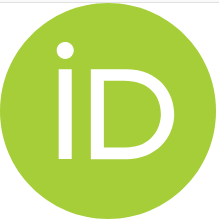}}}

\author{Tim Kokkeler\orcid{0000-0001-8681-3376}}
\email{tim.kokkeler@dipc.org}
\affiliation{Donostia International Physics Center (DIPC), 20018 Donostia--San Sebasti\'an, Spain}
\affiliation{University of Twente, 7522 NB Enschede, The Netherlands}

\author{Alexander Golubov\orcid{0000-0001-5085-5195}}
\affiliation{University of Twente, 7522 NB Enschede, The Netherlands}
\author{Sebastian Bergeret\orcid{0000-0001-6007-4878}}
\affiliation{Centro de F\'isica de Materiales (CFM-MPC) Centro Mixto CSIC-UPV/EHU, E-20018 Donostia-San Sebasti\'an,  Spain}
\affiliation{Donostia International Physics Center (DIPC), 20018 Donostia--San Sebasti\'an, Spain}
\author{Yukio Tanaka}
\affiliation{Deparment of Applied Physics, Nagoya University, 464-8603 Nagoya, Japan\\
}
%\begin{document}
\begin{abstract}
    In non-centrosymmetric superconductors the pair potential has both even-parity singlet and odd-parity triplet components. If time-reversal symmetry is broken, the superconducting phase of these components is not the same, for example in anapole superconductors. In this paper it is shown that breaking time-reversal symmetry by a phase difference between the two components significantly alters both the density of states and the conductance in s+helical p-wave superconductors. The density of states and conductance in s+chiral p-wave superconductors are less influenced by adding a phase difference because time reversal symmetry is already broken in the s+p-wave superconductor. The Tanaka-Nazarov boundary conditions are extended to 3D superconductors, allowing to investigate a greater variety of superconductors, such as B-W superconductors, in which the direction of the d-vector is parallel to the direction of momentum. The results are important for the determination of pair potentials in potentially time-reversal symmetry broken non-centrosymmetric superconductors.
\end{abstract}
\maketitle
%\onecolumngrid\
\section{Introduction}
\begin{figure}
    \centering
    \includegraphics[width = 8.6cm]{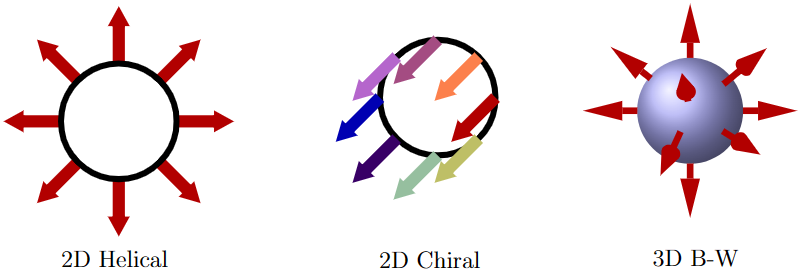}
    \caption{The three different types of p-wave pair potentials used in this paper. The d-vector is defined by its direction $\vec{\hat{d}}$ and its phase $\psi$, which is illustrated using colour. For both the 2D helical and the 3D B-W superconductor the phase the d-vector is real, $\psi$ is constant, but the direction of the d-vector is momentum dependent. On the other hand, for the 2D chiral superconductor the phase depends on momentum, while the direction of the d-vector is constant over the Fermi surface. In each case the magnitude of the gap is isotropic.}
    \label{fig:TypesOfPwave}
\end{figure}
Ever since the discovery of high temperature superconductors, much attention has been paid to unconventional superconductors \cite{sigrist1991phenomenological,schnyder2008classification,kallin2009sr2ruo4,maeno2011evaluation,sigrist2005review},  with for example triplet \cite{mackenzie2017even,kallin2016chiral,linder2019odd,balian1963superconductivity,sigrist1991phenomenological,fu2008superconducting,chiu2021observation,bauer2004heavy,linder2015superconducting,suh2020stabilizing,aoki2019review, saxena2000superconductivity,aoki2001coexistence,hardy2005p,huy2007super,ran2019nearly,yang2021spin,tanaka2004anomalous,tanaka2005theory} or odd-frequency \cite{berezinskii1974new,balatsky1992new,belitz1992even,abrahams1995properties,kirkpatrick1991disorder,coleman1997reflections,linder2019odd,cayao2020odd,gentile2011odd,tanaka2007theory,tanaka2011symmetry,tsintzis2019odd,kuzmanovski2020suppression} pairing. %, which may exist either intrinsically or proximized in junctions by for example a ferromagnet.
Historically, most attention has been paid to superconductors in which time reversal symmetry and inversion symmetry are not broken. In such superconductors the pair potential is either even-parity or odd-parity. However, if inversion symmetry is broken, a type of superconductivity can emerge that is neither even-parity or odd-parity \cite{bauer2012non}. Such superconductivity, with both singlet and triplet components appears in materials whose crystal structure breaks inversion symmetry 
\cite{bauer2004heavy,amano2004superconductivity,akazawa2004pressure,togano2004Superconductivity,tateiwa2005novel,kimura2005pressure,sugitani2006pressure,honda2010pressure,SETTAI2007844,Bauer2010Unconventional, xie2020captas}. In certain materials even the mixing parameter, the ratio between the singlet and triplet components, can be varied using electron irradiation \cite{ishihara2021tuning}.

Moreover, there exists several unconventional superconductors, including possibly $\text{Sr}_{2}\text{RuO}_{4}$, in which time reversal symmetry is  broken \cite{wenger1993dwave,rokhsar1993pairing,covington1997observation,sigrist1991phenomenological,sigrist1991phenomenology,laughlin1994tunneling,laughlin1998magnetic,hillier2009evidence,wakatsuki2017nonreciprocal,kivelson2020proposal,suh2020stabilizing,xia2006high,luke1998time,clepkens2021shadowed,sigrist1998time,ghosh2020recent,ghosh2022time,lee2009pairing,farhang2023revealing,ajeesh2023fate,shang2020simultaneous,grinenko2021split,willa2021inhomogeneous,maisuradze2010evidence,roising2022heat,movshovich1998low,balatsky1998spontaneous,belyavsky2012chiral,black2014chiral}. For example, in chiral superconductors \cite{kallin2016chiral}, time-reversal symmetry is broken in the bulk. Next to this, time reversal symmetry has been predicted to be spontaneously broken near the surface of d-wave superconductors \cite{tanaka2001influence,sigrist1995fractional,palumbo1990magnetic,kuboki1996proximity,tanuma1999quasiparticle,tanuma2001tunneling,fogelstrom1997tunneling,krupke1999anisotropy,matsumoto1995coexistence}.
Similarly, in so-called anapole superconductors there exists a nonzero phase difference between the singlet and the triplet components \cite{kanasugi2022anapole,kitamura2022quantum,chazono2022piezoelectric,goswami2014axionic,mockli2019magnetic}. In these superconductors, both time-reversal symmetry and inversion symmetry is broken while the product of time reversal and inversion symmetry is preserved, thereby providing an analogy to axion electrodynamics \cite{wilczek1987two,qi2013axion}. An example of such superconductor is the is+p-wave superconductor.

The proximity effect of non-centrosymmetric or time reversal symmetry broken superconductors has been studied in detail in several limits \cite{iniotakis2007andreev,eschrig2010theoretical,annunziata2012proximity,rahnavard2014magnetic, mishra2021effects,ikegaya2021proposal,daido2017majorana,chiu2023tuning}.
Recently a theory has been developed to calculate the proximity effect of non-centrosymmetric superconductors in dirty normal metals using the Keldysh Usadel formalism \cite{tanaka2022theory,kokkeler2023spin,kokkeler2023anisotropic}. Those works focus on time-reversal symmetric s+helical p-wave superconductors and on s+chiral superconductors in which there is no phase difference between the singlet and triplet components of the mode of normal incidence. In this work, we use this theory to explore the density of states, pair amplitudes, and conductance in the presence of an arbitrary phase difference between the s-wave and p-wave components of the pair potential in the superconductor. We will refer to such superconductors as (i)s+p-wave superconductors. We calculate the conductance in SNN junctions with (i)s+p-wave superconductors.  We show that the phase difference has a large influence on the proximity effect induced by (i)s+helical p-wave, and show that in the absence of time-reversal symmetry the density of states, pair amplitudes, and conductance are significantly altered. Notably, the quantization of the zero energy density of states and the zero bias conductance in s+helical p-junctions disappears in the presence of a phase difference between the singlet and triplet components. %because there is no topological protection in the absence of time-reversal symmetry. 
Next to this, we show that for (i)s+chiral p-wave superconductors, for which time reversal symmetry is broken by the p-wave component of the pair potential even in the absence of an s-wave component, the phase difference between the singlet and triplet correlations has a much smaller influence on the proximity effect. Our results highlight the importance of confirming the presence or absence of time-reversal symmetry in non-centrosymmetric superconductors.

Since the proposed anapole superconductors are three-dimensional \cite{kanasugi2022anapole}, we extend the Tanaka-Nazarov boundary condition to 3D unconventional superconductors and study the Balian-Werthamer (B-W) phase that was first found in Helium \cite{buchholtz1981identification,buchholtz1986fermi,higashitani2009proximity}. The B-W phase is the natural three-dimensional generalization of the 2D helical p-wave phase, the d-vector is real and parallel to the direction of momentum and the magnitude of the gap is isotropic.
We show that contrary to junctions with 2D p-wave superconductors, junctions with 3D p-wave superconductors may have a zero bias conductance dip instead of a peak in short diffusive junctions. For long junctions, there exists a sharp peak with a width of the order of the Thouless energy due to coherent Andreev reflection. %We show that the 

\section{Theory}
\begin{figure}
    \centering
    \includegraphics[width = 8.6cm]{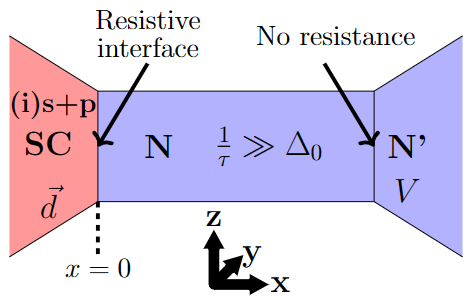}
    \caption{The setup used in our calculations. The junction consists of a dirty ($1/\tau\gg\Delta_{0}$) normal metal bar sandwiched between a superconducting electrode and a normal metal. The superconductor has both singlet s-wave correlations and triplet p-wave correlations characterized by the d-vector $\vec{d}$. A voltage $V$ is applied to the normal metal, while the superconducting electrode is grounded. The interface with the normal metal electrode has no boundary resistance ($\gamma_{B} =0$). The interface with the superconductor is resistive, and Tanaka-Nazarov boundary conditions are used.}
    \label{fig:SNNSetup}
\end{figure}

We consider the SNN-junction shown in Fig. \ref{fig:SNNSetup}, consisting of a dirty normal metal bar, with a scattering rate $1/\tau$ much larger than any other relevant energy scale, sandwiched between a normal metal electrode to which a voltage may be applied, and an (i)s+p-wave superconductor with arbitrary phase difference between the singlet and triplet components:
\begin{align}
    \hat{\Delta}(\phi) = \Delta_{0}\left(e^{i\frac{\pi}{2}\chi_{t}} \frac{1}{\sqrt{r^{2}+1}}+\frac{r}{\sqrt{r^{2}+1}}\vec{d}(\phi)\cdot\vec{\sigma}\right),\label{eq:PairPotential}
\end{align}
where $\Delta_{0}$ is a real scalar, $\vec{d}(\phi)$ is the d-vector, an angle dependent vector that is conventionally used to describe the spin-dependence of the pair potential,  $\phi$ is the angle between the direction of momentum and the normal to the S/N surface in Fig. \ref{fig:SNNSetup}, $\vec{\sigma}$ is the vector of Pauli matrices in spin space, $r$ is the mixing parameter between the even-parity singlet s-wave component and the odd-parity triplet p-wave component of the pair potential and $\chi_{t}$ is the phase difference between the singlet and triplet components. The singlet component of the pair potential will be referred to as $\Delta_{s}$, the triplet as $\Delta_{p}$.
For $r = 0$ the expression reduces to conventional s-wave superconductivity, as $r\xrightarrow{}\infty$ a p-wave superconductor is obtained. The d-vector is different for different types of p-wave superconductivity. Because p-wave superconductors are odd-parity, it necessarily satisfies $\vec{d}(\phi+\pi) = -\vec{d}(\phi)$.
%We first focus on (i)s+helical p-wave superconductors, that is, with d-vector
%\begin{align}
%    \vec{d}(\phi) = (\cos{\phi},\sin{\phi},0).
%\end{align}
We consider the density of states and pair amplitude in the normal metal bar at the interface with the superconducting electrode, and the conductance through the junction when applying a voltage to the normal metal, see Fig. \ref{fig:SNNSetup}.
\begin{comment}
We assume that the F (N) is dirty, so that the Green's function in the bar is almost isotropic. Hence, the Usadel formalism can be used:
\begin{align}
    D\nabla(\check{G}\nabla\check{G}) = [(iE+\vec{h}\cdot\vec{\sigma})\tau_{3},\check{G}],
\end{align}
%where $\check{G}$ is the Green's function, $D$ is the diffusion constant in the ferromagnet, $E$ is energy, $\vec{h}$ is the exchange field, $\vec{\sigma}$ is the vector of Pauli matrices in spin space and $\tau_{3}$ is the third Pauli matrix in Nambu space.
We supply the Usadel equation with the following boundary conditions. 
For the boundary between the bar and the superconductor we use the Tanaka-Nazarov boundary conditions, which are suited to the interface between unconventional superconductors and dirty materials:
\begin{align}
    \check{G}\nabla \check{G} &= \check{I}\\
    \check{I} &=\int (1+T_{1}^{2}+T_{1}(CG+GC))^{-1}T_{1}(CG-GC)\cos\phi d\phi
\end{align}
\end{comment} 
We assume that the normal metal bar is in the dirty limit, that is, the scattering rate is high and the Green's function is almost isotropic. This allows us to use the Usadel equation to describe the N \cite{belzig1999quasiclassical,chandrasekhar2008proximity}. Moreover, we assume that the bar is either very wide or very narrow compared to the thermal diffusion length in the directions perpendicular to the transport direction, so that an effectively one-dimensional model may be used:
\begin{align}
    D\partial_{x}(\check{G}\partial_{x} \check{G}) = [iE\tau_{3},\check{G}],
\end{align}
where $D$ is the diffusion constant of the normal metal bar, $\check{G}$ is the isotropic component of the Green's function in Keldysh-Nambu-spin space and $E$ is energy.
The contact between the bar and the normal metal electrode at $x  = L$ is assumed to be very good, $\gamma_{BN} = 0$, so that the Green's function is continuous at this interface:
\begin{align}
    \check{G}(x = L) = \check{G}_{N},
\end{align}
where $\check{G}_{N}$ is Green's function in the normal metal electrode. It is equal to the Green's function in the bulk of a normal metal, with retarded part given by $\check{G}_{N}^{R} = \tau_{3}$, and distribution functions $f_{L,T} = \frac{1}{2}\left(\tanh\frac{E+eV}{2kT}\pm\tanh\frac{E-eV}{2kT}\right)$. At the interface with the superconductor ($x = 0$) we use the Tanaka-Nazarov boundary conditions \cite{tanaka2003circuit,tanaka2004theory}, the extension of Nazarov's boundary conditions \cite{nazarov1999novel} to junctions with unconventional superconductors:
\begin{equation}\label{eq:Tanaka-Nazarov}
    \check{G}\nabla \check{G}(x=0) = \frac{1}{\gamma_{BS} L}\langle \check{S}(\phi)\rangle \; ,
\end{equation}
where \cite{tanaka2022theory}:
\begin{align}
    \check{S}(\phi) &= \Tilde{T}(1+T_{1}^{2}+T_{1}(\check{C}\check{G}+\check{G}\check{C}))^{-1}(\check{C}\check{G}-\check{G}\check{C})\; ,\\
    \check{C} &=\check{H}_{+}^{-1}(\check{\mathbf{1}}-\check{H}_{-})\label{eq:Cdef}\;,\\
    \check{H}_{+}&=\frac{1}{2}(\check{G}_{S}(\phi)+\check{G}_{S}(\pi-\phi))\label{eq:Hplus}\; ,\\
    \check{H}_{-}&=\frac{1}{2}(\check{G}_{S}(\phi)-\check{G}_{S}(\pi-\phi))\label{eq:Hmin}\;.
\end{align}
Here $\check{G}_{S}(\phi)$ is the bulk Green's function of an (i)s+p-wave superconductor given. For the superconductors discussed in this paper the magnitude of the triplet component is constant over the Fermi surface, hence $\vec{d}(\phi) $ can be written as $\vec{\hat{d}}(\phi)e^{i\psi(\phi)}$, where $\vec{\hat{d}}(\phi)$ is a unit vector and $\psi(\phi)$ is a real phase. Specifically, for helical superconductors $\psi(\phi) = 0$ and $\vec{\hat{d}}(\phi) = (\cos\phi,\sin\phi,0)$, while for chiral superconductors $\psi(\phi) = \phi$ and $\vec{\hat{d}}(\phi) = (0,0,1)$. 
Therefore, using the basis $(\psi_{\uparrow},\psi_{\downarrow},\psi^{\dagger}_{\downarrow},-\psi^{\dagger}_{\uparrow})$, the bulk Green's function $\check{G}_{S}(\phi)$ is given by \cite{kokkeler2023spin}
\begin{align}
    \check{G}_{s}(\phi) &= \frac{1}{2}(1+\vec{\hat{d}}(\phi)\cdot\vec{\sigma})\otimes\frac{1}{\sqrt{E^{2}-|\Delta_{+}|^{2}}}\begin{bmatrix}
        E&\Delta_{+}\\-\Delta_{+}^{*}&-E
    \end{bmatrix}\nonumber\\&+\frac{1}{2}(1-\vec{\hat{d}}(\phi)\cdot\vec{\sigma})\otimes\frac{1}{\sqrt{E^{2}-|\Delta_{-}|^{2}}}\begin{bmatrix}
        E&\Delta_{-}\\-\Delta_{-}^{*}&-E
    \end{bmatrix}\; ,\nonumber\\
    \Delta_{\pm} &= \frac{e^{i\frac{\pi}{2}\chi_{t}}\pm re^{i\psi(\phi)}}{\sqrt{r^{2}+1}}\;.
\end{align}
The brackets $\langle\cdot\rangle$ indicate angular averaging over all modes that pass through the interface, the symbol $\otimes$ is used to denote a Kronecker product, $\gamma_{BS} = R_{B}/R_{d}$ is the ratio between the boundary resistance and the normal state resistance of the dirty normal metal bar,  $T_{1} = \Tilde{T}/(2-\Tilde{T}+2\sqrt{1-\Tilde{T}})$, and $\Tilde{T}$ is the interface transparency given by \cite{tanaka2005theory}:
\begin{align}\label{eq:zdef}
\Tilde{T}(\phi) = \frac{\cos^{2}\phi}{\cos^{2}{\phi}+z^{2}}\; ,
\end{align} 
where $z$ is the Blonder-Tinkham-Klapwijk (BTK) parameter. We do not take into account the Fermi surface mismatch, assuming that the magnitude of the Fermi momentum is of similar magnitude in the superconductor and normal metal.

The equations for the retarded part $\check{G}^{R}$ in Keldysh space are solved numerically. 
The advanced Green's function $\check{G}^{A}$ is directly related to the retarded part via $\check{G}^{A} = -\tau_{3}(\check{G}^{R})^{\dagger}\tau_{3}$ \cite{belzig1999quasiclassical,chandrasekhar2008proximity}. Based on the solutions for the retarded and advanced components, the Keldysh component can be found numerically using the distribution function $\check{h} = \hat{f}_{L}\otimes\mathbf{1}_{\tau}+\hat{f}_{T}\otimes\tau_{3}$ for $\check{G}^{K} = \check{G}^{R}\check{h}-\check{h}\check{G}^{A}$ \cite{belzig1999quasiclassical,chandrasekhar2008proximity},
where $\mathbf{1}_{\tau}$ is the identity matrix in Nambu space, $\tau_{3}$ is the third Pauli matrix in Nambu space and $\hat{f}_{L}$ and $\hat{f}_{T}$ are the so-called longitudinal and transverse components of the  distribution function \cite{schmid1975linearized,heikkila2019thermal} to be determined.

We consider the density of states and pair amplitudes at the interface with the superconductor, given by
\begin{align}
    \rho &= \text{Tr}\Big((\mathbf{1}_{\sigma}\otimes\tau_{3})\check{G}^{R}(x = 0)\Big)\\
    F^{R}_{s1,2} &= \text{Tr}\Big((\mathbf{1}_{\sigma}\otimes\tau_{1,2})\check{G}^{R}(x =0)\Big)\\
    F^{R}_{t1,2} &= \text{Tr}\Big((\vec{\hat{d}}\cdot\vec{\sigma} \otimes \tau_{1,2})\check{G}^{R}(x =0)\Big),
\end{align}
where $\tau_{1,2,3}$ are the Pauli matrices in Nambu space and $\sigma_{1,2,3}$ the Pauli matrices in spin space, while $\mathbf{1}_{\sigma}$ is the identity matrix in spin space.
Due to  the Pauli principle, the singlet correlations are even-frequency correlations on the Matsuraba axis, thus satisfying $F^{R}_{s1,2}(-E) = \left(F^{R}_{s1,2}(E)\right)^{*}$, while the triplet correlations are odd-frequency correlations, therefore satisfying $F^{R}_{t1,2}(-E) = -\left(F^{R}_{t1,2}(E)\right)^{*}$. %\TK{I do not think it is good to put the following the final version, but for internal notes: these relations can be found as follows. For even-frequency correlations $F^{R}(-E) = \lim_{\omega\xrightarrow{}-E+i\delta}F(\omega) = \lim_{\omega\xrightarrow{}-E+i\delta}F(-\omega)=\lim_{\omega\xrightarrow{}E-i\delta}F(\omega) = F^{A}(E) = (F^{R}(E))^*$, where the last equality follows from $G^{R} = -\tau_{3}(G^{R})^{\dagger}\tau_{3}$. Similarly, for odd-frequency amplitudes $F^{R}(-E) = \lim_{\omega\xrightarrow{}-E+i\delta}F(\omega) = \lim_{\omega\xrightarrow{}-E+i\delta}F(-\omega)=-\lim_{\omega\xrightarrow{}E-i\delta}F(\omega) = -F^{A}(E) = -(F^{R}(E))^*$.}

Using the Keldysh component the conductance can be calculated as
\begin{align}
\sigma  &= \frac{\partial I}{\partial V}\nonumber\\
I&=\frac{\sigma_{N}}{16e}\int_{-\infty}^{\infty} \mathrm{d}E\text{Tr}\left\{(\mathbf{1}_{\sigma}\otimes\tau_{3})(\check{G}\nabla \check{G})^{K}\right\}\,
\end{align}
where $\sigma_{N}$ is the normal state conductance of the bar and $(\check{G}\nabla \check{G})^{K}$ is the Keldysh component of $\check{G}\nabla \check{G}$.

%This implies that in the clean limit, the BTK model predicts that for $\chi_{t}\neq 1$ there are two peaks at $eV/\Delta_{0} = \sqrt{1\pm\frac{2r\cos{\frac{\pi}{2}\xi_{t}}}{1+r^{2}}}$, which combine into a single peak for $eV = \Delta_{0}$.
\section{Helical p-wave superconductors}
In this section, we present the result for (i)s+helical p-wave superconductors. For helical p-wave superconductors the d-vector is given by $\vec{d}(\phi) = (\cos\phi,\sin\phi,0)$.
For $\chi_{t} = 0$, there is no time-reversal symmetry breaking and this model reduces to the model studied in \cite{tanaka2022theory,kokkeler2023spin,kokkeler2023anisotropic}. Another special case is $\chi_{t} = 1$. In that case the quantities
\begin{align}
    \Delta_{\pm} = \frac{e^{i\frac{\pi}{2}\chi_{t}}\pm r}{\sqrt{1+r^{2}}}\Delta_{0}
\end{align}
both have magnitude $\Delta_{0}$, irrespective of $r$ or angle $\phi$. This implies that the double peak structure obtained for $\chi_{t} = 0$ is not 
to be expected for $\chi_{t} = 1$.

%Moreover, the quantization of the zero bias conductance, which can be understood in terms of $\Delta_{\pm}/|\Delta_{\pm}|$ \cite{ikegaya2016quantization}, need not be present in the absence of time-reversal symmetry.
The dependence of $|\Delta_{\pm}|$ on $\chi_{t}$, $r$ and  $\phi$ is described by:
\begin{align}
    |\Delta_{\pm}|^{2} = (1\pm\frac{2r\cos{\frac{\pi}{2}\chi_{t}}}{1+r^{2}})\Delta_{0}^{2}.
\end{align}
In the next section, we discuss the dependence of the density of states, pair amplitudes, and conductance on $\chi_{t}$ for various ratios of the s-wave and p-wave components of the pair potential while setting the other parameters equal to the values used in \cite{kokkeler2023anisotropic} for the case $\chi_{t} = 0$, that is, $\gamma_{B} =2$, $z = 0.75$ and $\Delta_{0}/E_{\text{Th}} = 50$.
\subsection{Density of states and pair amplitudes}
\begin{figure*}
    \centering
    {\hspace*{-2em}\includegraphics[width = 8.6cm]{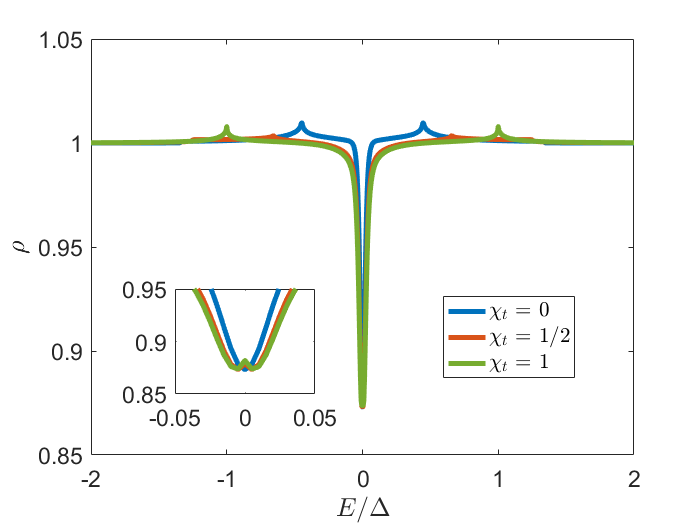}}
    \hfill
    {\hspace*{-2em}\includegraphics[width = 8.6cm]{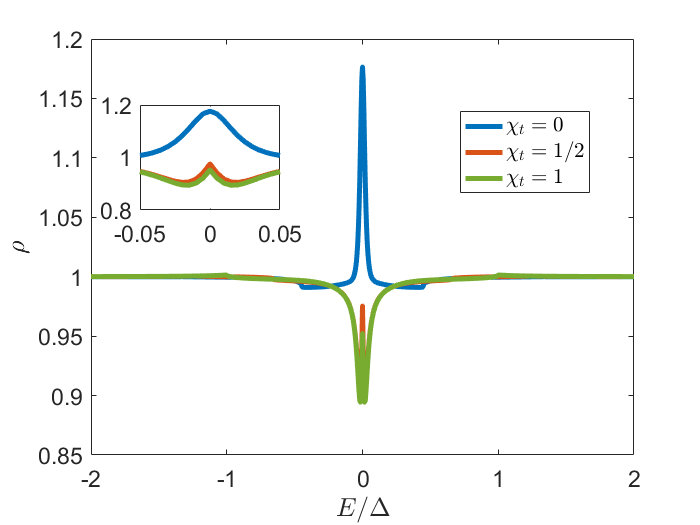}}
    \hfill
    {\hspace*{-1.5em}\includegraphics[width = 8.6cm]{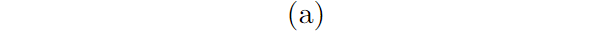}}
    \hfill
    {\hspace*{-2em}\includegraphics[width = 8.6cm]{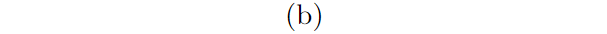}}
    \caption{The local density of states for $e^{i\chi_{t}\frac{\pi}{2}}s+p$-wave superconductors for s-wave dominant ($r = 0.5$) and p-wave dominant $(r = 2)$ superconductors. Other parameters are set to $\gamma_{B} = 2$, $z = 0.75$, $E_{\text{Th}}/\Delta_{0} = 0.02$.}\label{fig:LDOS2D}
\end{figure*}
From the retarded part of the Green's function $G^{R}$, the local density of states at the SN-interface ($x = 0$) can be extracted. The local density of states is shown in Fig. \ref{fig:LDOS2D}. 
 For s-wave dominant superconductors, Fig. \ref{fig:LDOS2D}(a), the zero energy density of states is almost independent of $\chi_{t}$, the phase difference between the s-wave and p-wave components of the pair potential only influences the details of the density of states. Most notably, the most pronounced peak shifts from $E = \Delta_{-}$ to $E = \Delta_{0}$ with increasing $\chi_{t}$. On the other hand, for p-wave dominant superconductors the zero energy density of states is influenced by $\chi_{t}$, as shown in Fig. \ref{fig:LDOS2D}(b). For $\chi_{t} = 0$ there is a large zero energy peak. If $\chi_{t}$ is nonzero the zero energy peak is highly suppressed, there are both a broad dip and a sharp peak centered at zero energy. In contrast to junctions with s+helical p-wave superconductors with $\chi_{t} = 0$ the zero energy density of states is not guaranteed to be higher than the normal density of states.
 \begin{figure*}
    \centering
    {\hspace*{-2em}\includegraphics[width = 4.3cm]{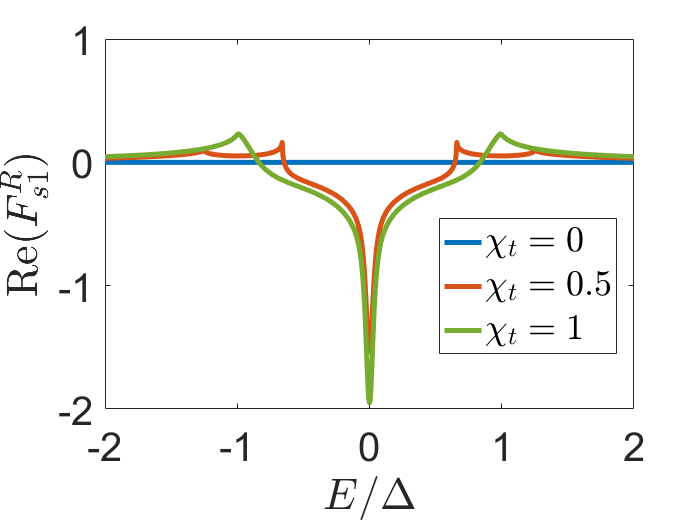}}
    \hfill
    {\hspace*{-2em}\includegraphics[width = 4.3cm]{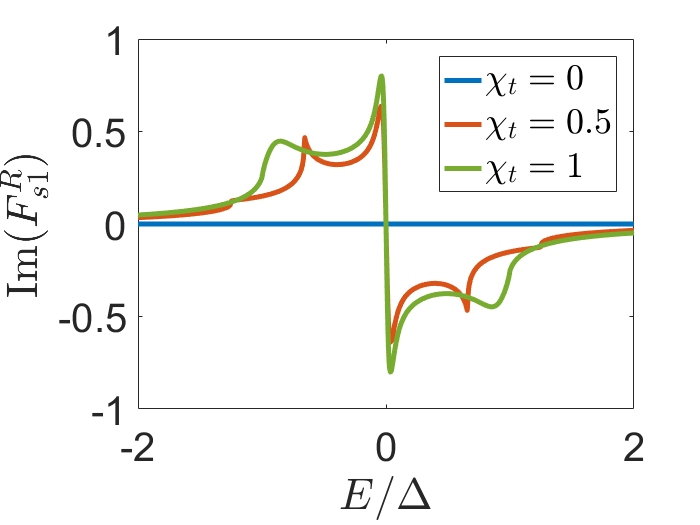}}
    \hfill
    {\hspace*{-2em}\includegraphics[width = 4.3cm]{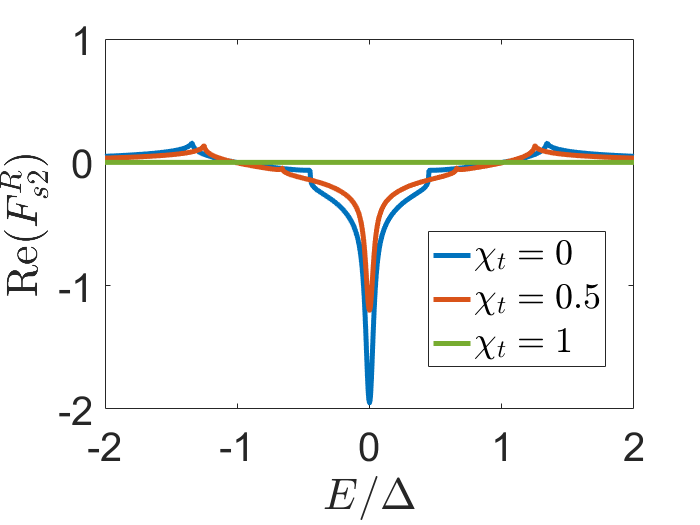}}
    \hfill
    {\hspace*{-2em}\includegraphics[width = 4.3cm]{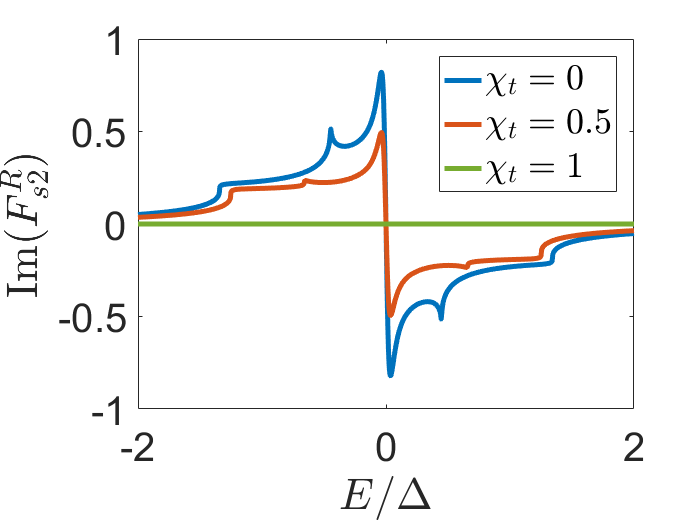}}
    \hfill
    {\hspace*{-2em}\includegraphics[width = 4.3cm]{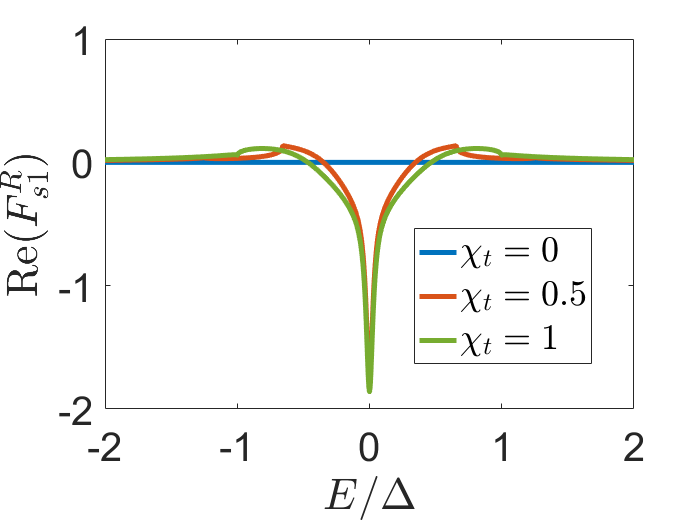}}
    \hfill
    {\hspace*{-2em}\includegraphics[width = 4.3cm]{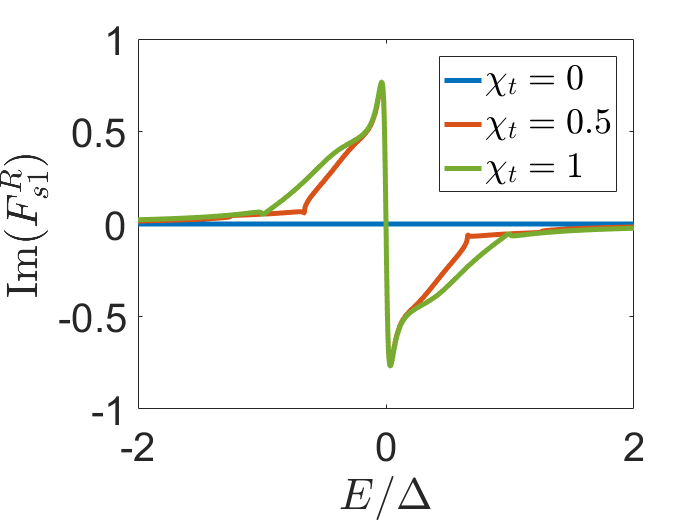}}
    \hfill
    {\hspace*{-2em}\includegraphics[width = 4.3cm]{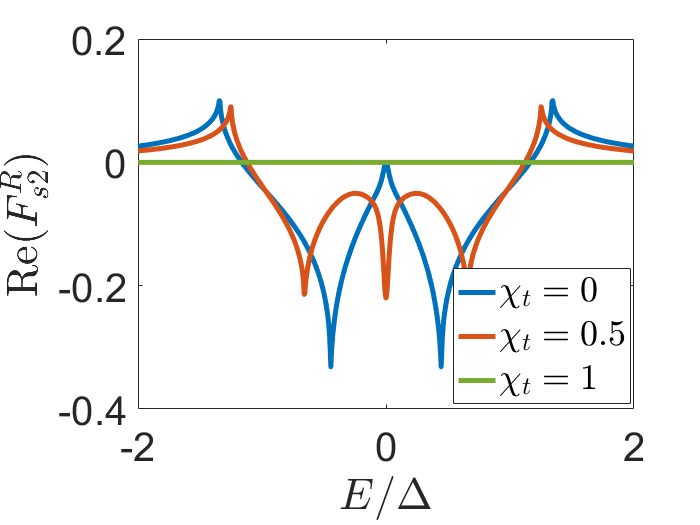}}
    \hfill
    {\hspace*{-2em}\includegraphics[width = 4.3cm]{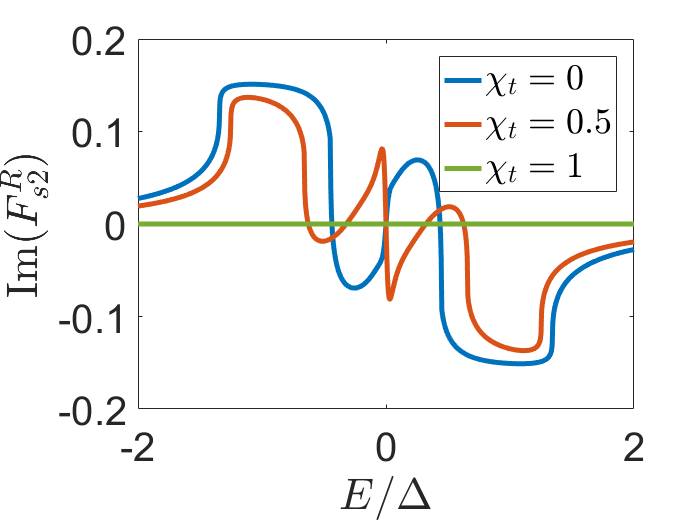}}
    \caption{The singlet pair amplitudes for $e^{i\chi_{t}\frac{\pi}{2}}s$+helical p-wave superconductors for s-wave dominant superconductors with $r = 0.5$ (top row) and p-wave dominant superconductors with $r = 2$ (bottom row). Other parameters are set to $\gamma_{B} = 2$, $z = 0.75$, $E_{\text{Th}}/\Delta_{0} = 0.02$. }\label{fig:PAS2D}
\end{figure*}
\begin{figure*}
    \centering
    {\hspace*{-2em}\includegraphics[width = 4.3cm]{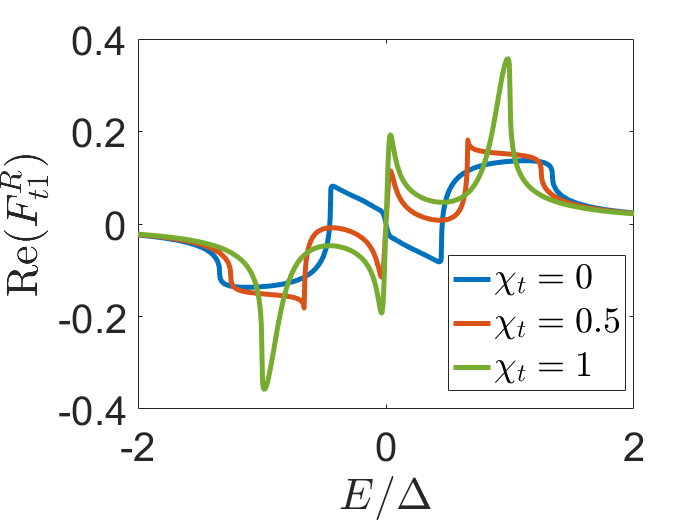}}
    \hfill
    {\hspace*{-2em}\includegraphics[width = 4.3cm]{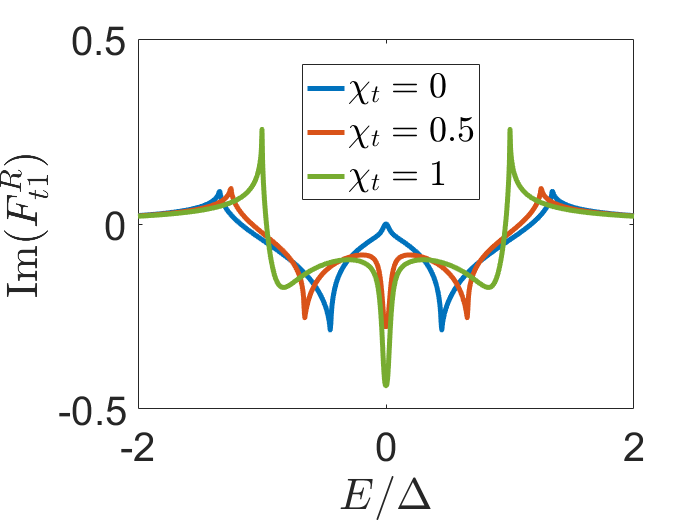}}
    \hfill
    {\hspace*{-2em}\includegraphics[width = 4.3cm]{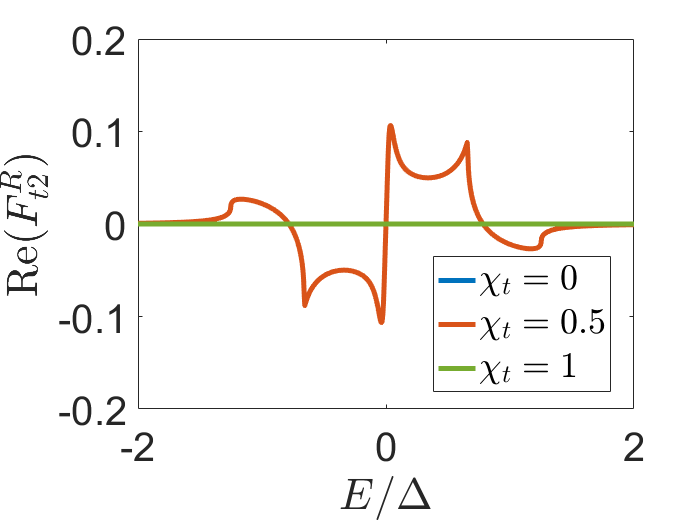}}
    \hfill
    {\hspace*{-2em}\includegraphics[width = 4.3cm]{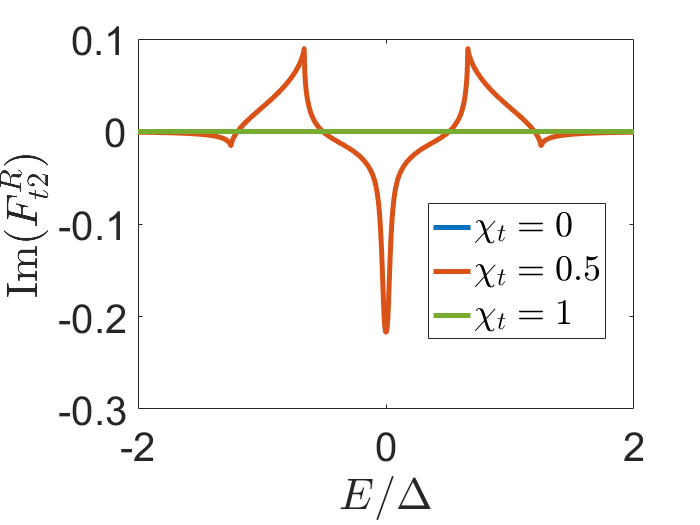}}
    \hfill
    {\hspace*{-2em}\includegraphics[width = 4.3cm]{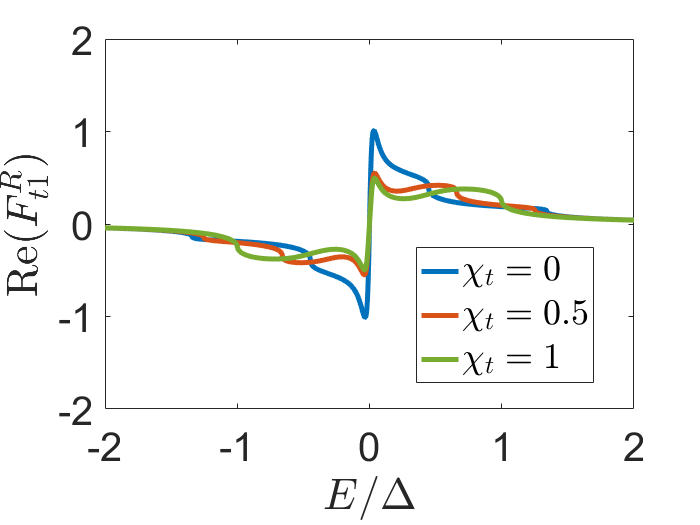}}
    \hfill
    {\hspace*{-2em}\includegraphics[width = 4.3cm]{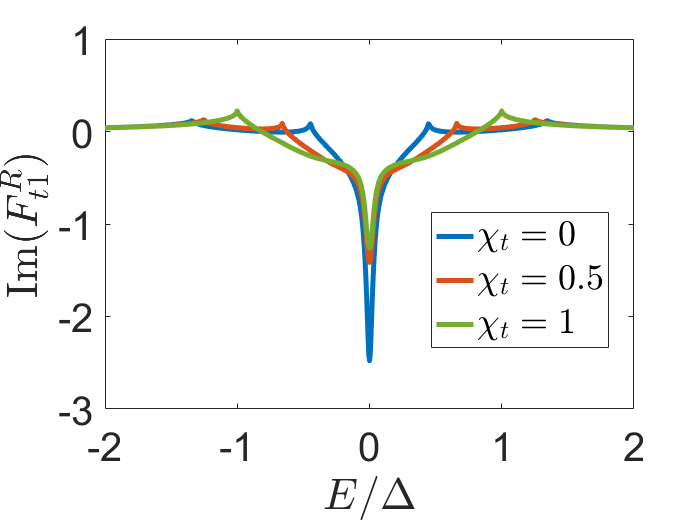}}
    \hfill
    {\hspace*{-2em}\includegraphics[width = 4.3cm]{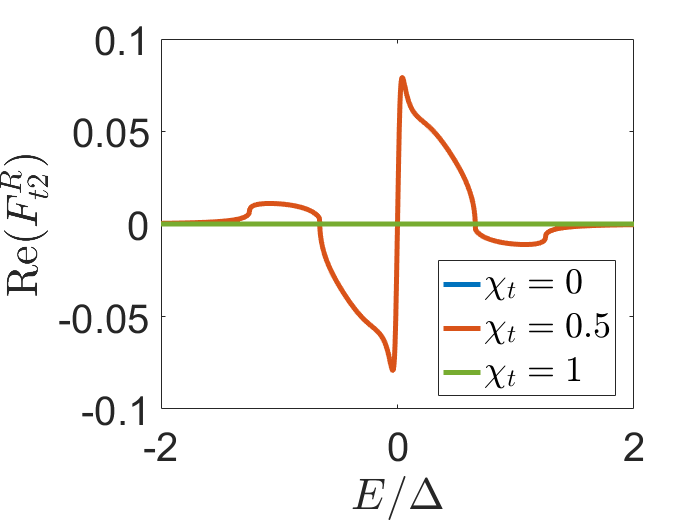}}
    \hfill
    {\hspace*{-2em}\includegraphics[width = 4.3cm]{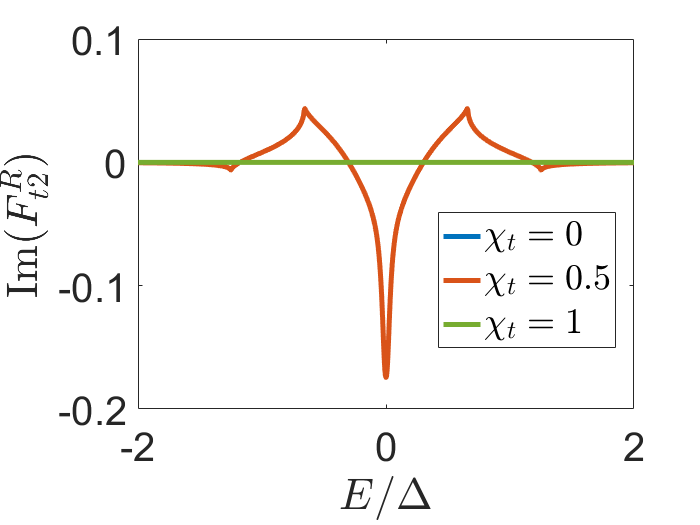}}
    \caption{The triplet pair amplitudes for $e^{i\chi_{t}\frac{\pi}{2}}s$+helical p-wave superconductors for s-wave dominant superconductors with $r = 0.5$ (top row) and p-wave dominant superconductors with $r = 2$ (bottom row). Other parameters are set to $\gamma_{B} = 2$, $z = 0.75$, $E_{\text{Th}}/\Delta_{0} = 0.02$. }\label{fig:PAT2D}
\end{figure*}
 \begin{comment}
\begin{figure*}
    \centering
    {\hspace*{-2em}\includegraphics[width = 8.6cm]{figures/SingletPairAmplitudeR0p5h0.png}}
    \hfill
    {\hspace*{-2em}\includegraphics[width = 8.6cm]{figures/SingletPairAmplitudeR2h0.png}}
    \hfill
    {\hspace*{-1.5em}\includegraphics[width = 8.6cm]{figures/A.png}}
    \hfill
    {\hspace*{-2em}\includegraphics[width = 8.6cm]{figures/B.png}}
    \caption{The singlet pair amplitudes for $e^{i\chi_{t}\frac{\pi}{2}}s+p$-wave superconductors for s-wave dominant ($r = 0.5$) and p-wave dominant $(r = 2)$ superconductors. Other parameters are set to $\gamma_{B} = 2$, $z = 0.75$, $E_{\text{Th}}/\Delta_{0} = 0.02$. }\label{fig:PAS2D}
\end{figure*}
\begin{figure*}
    \centering
    {\hspace*{-2em}\includegraphics[width = 8.6cm]{figures/TripletZPAR0p5h0.png}}
    \hfill
    {\hspace*{-2em}\includegraphics[width = 8.6cm]{figures/TripletZPAR2h0.png}}
    \hfill
    {\hspace*{-1.5em}\includegraphics[width = 8.6cm]{figures/A.png}}
    \hfill
    {\hspace*{-2em}\includegraphics[width = 8.6cm]{figures/B.png}}
    \caption{The triplet pair amplitudes for $e^{i\chi_{t}\frac{\pi}{2}}$s+helical p-wave superconductors for s-wave dominant ($r = 0.5$) and p-wave dominant $(r = 2)$ superconductors. Other parameters are set to $\gamma_{B} = 2$, $z = 0.75$, $E_{\text{Th}}/\Delta_{0} = 0.02$.}\label{fig:PAT2D}
\end{figure*}
\end{comment}

The singlet and triplet pair amplitudes are shown respectively in Figs. \ref{fig:PAS2D} and \ref{fig:PAT2D}. 
The phase of the induced correlations depends on $\chi_{t}$. For $\chi_{t} = 0$, that is, for s+p-wave superconductors, only $F_{s2}$ and $F_{t1}$ can be nonzero. On the other hand, for $\chi_{t} =1$, that is, for is+p-wave superconductors, only $F_{s1}$ and $F_{t1}$ are nonzero. For $0<\chi_{t}<1$ all components of the pair amplitude can be nonzero and the phase of the correlations depends on energy.
%Next to this, the pair amplitudes obey the symmetries $F_{s1,2}(-E) = F_{s1,2}(E)^{*}$ and $F_{t1,2} = -F_{t1,2}(-E)^{*}$ as discussed in the previous section. This implies that at $E = 0$ only $\text{Re}(F_{s1,2})$ and $\text{Im}(F_{t1,2})$ can be nonzero.

For s-wave dominant superconductors, the real part of the singlet pair amplitudes, see Fig. \ref{fig:PAS2D}(top), is  maximized at zero energy, regardless of $\chi_{t}$. %The phase difference $\chi_{t}$ has a large influence on the phase of the induced correlations for $\chi_{t} = 0$ only the $\tau_{2}$-component is nonzero, for $\chi_{t} = 1$ only the $\tau_{1}$-component is nonzero. For $\chi_{t} = 0.5$ both are present. The ratio between depends on energy, indicating that the induced singlet pairs have an energy dependent phase.
On the other hand, for the p-wave dominant case, Fig. \ref{fig:PAS2D}(bottom),  the singlet pair amplitudes vanish at zero energy if $\chi_{t} = 0$. In contrast, for nonzero $\chi_{t}$ the real parts of singlet pair amplitude are maximized at zero energy. Also here, the phase depends strongly on $\chi_{t}$.

The triplet pair amplitudes depend more strongly on $\chi_{t}$ if the s-wave component of the pair potential is dominant. Indeed, for s-wave dominant junctions with $\chi_{t} = 0$, Fig. \ref{fig:PAT2D}(top), the triplet pair amplitudes vanish at zero energy. On the other hand, for $\chi_{t}\neq 0$ the imaginary parts of the triplet pair amplitudes are maximized at zero energy.
%, though this maximum is much smaller than the maximum for p-wave dominant superconductors. 
For p-wave dominant superconductors, the imaginary parts of the triplet pair amplitude are maximized at zero energy for all $\chi_{t}$, as shown in  Fig. \ref{fig:PAT2D}(bottom). An increase of the phase difference between the singlet and triplet components suppresses this maximum, but only slightly. Summarizing, a nonzero $\chi_{t}$ enhances the subdominant pair amplitudes at zero energy, while only moderately altering the magnitude of dominant pair amplitudes. The phase $\chi_{t}$ does in all cases have a large influence on the phase of the induced correlations.

The different behaviour of the local density of states and the pair amplitudes in the presence of a phase difference between singlet and triplet pair potentials can be explained using topology. The helical p-wave superconductor is a topological superconductor protected by time-reversal symmetry breaking \cite{sato2017topological,schnyder2008classification,leijnse2012introduction}. For s+helical p-wave superconductors, there is therefore a topological phase transition at $r = 1$. On the other hand, for is+p-wave superconductors time-reversal symmetry is broken and there is no topological phase transition, all properties change continuously from s-wave to p-wave as the mixing parameter $r$ is increased. For the zero energy density of states and pair amplitudes, this can be shown explicitly. The density of states and pair amplitude at zero energy are fully determined by $\frac{\Delta_{+}}{|\Delta_{+}|}$ and $\frac{\Delta_{-}}{|\Delta_{-}|}$ \cite{ikegaya2016quantization,kokkeler2023anisotropic}. If $\chi_{t} = 0$, both $\Delta_{+}$ and $\Delta_{-}$ are necessarily real, and therefore, using that $\Delta_{+}>0$ by definition, the zero energy density of states and pair potentials are fully determined by the sign of $\Delta_{-}$. However, if $\chi_{t}\neq 0$ the quantities $\frac{\Delta_{+}}{|\Delta_{+}|}$ and $\frac{\Delta_{-}}{|\Delta_{-}|}$ are both complex and  continuous functions of $r$. Therefore, at $E = 0$ the results are neither s-wave nor p-wave, but rather a mixture and both singlet and triplet correlations are present.
\subsection{Conductance}
\begin{figure*}
    \centering
    {\hspace*{-2em}\includegraphics[width = 8.6cm]{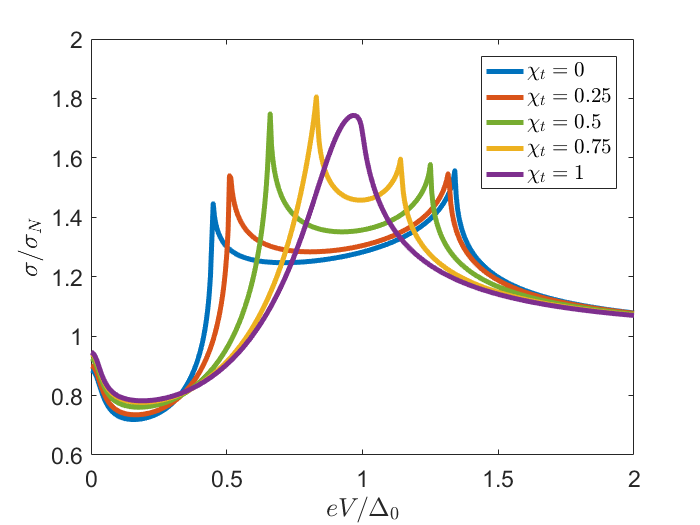}}
    \hfill
    {\hspace*{-2em}\includegraphics[width = 8.6cm]{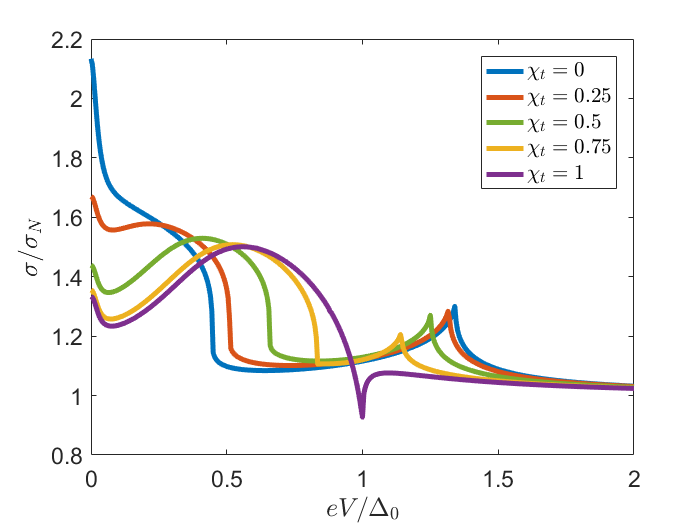}}
    \hfill
    {\hspace*{-1.5em}\includegraphics[width = 8.6cm]{figures/A.png}}
    \hfill
    {\hspace*{-2em}\includegraphics[width = 8.6cm]{figures/B.png}}
    \caption{The dependence of the conductance on the relative phase difference between the s-wave and helical p-wave components of the pair potential for $r = 0.5$ (left) and $r = 2$ (right). If there is no time reversal symmetry breaking ($\chi_t = 0$), the conductance has sharp features at $|\Delta_{\pm}| = |\Delta_{s}\pm\Delta_{p}|$. As the phase difference between the s-wave and p-wave components of the pair potential increases, these features come together, becoming one smooth feature for $\chi_{t} = 1$. Parameters are set to $\gamma_{B} = 2$, $z = 0.75$, $E_{\text{Th}}/\Delta_{0} = 0.02$.}
    \label{fig:TRBdeps}
\end{figure*}

The effect of a finite phase difference between the s-wave and the p-wave components of the pair potential on the conductance is illustrated in Fig. \ref{fig:TRBdeps} for an s-wave dominant superconductor with $r = 0.5$ (Fig. \ref{fig:TRBdeps}(a)) and a p-wave dominant superconductor with $r = 2$ (Fig. \ref{fig:TRBdeps}(b)). %For $r = 0.5$, as the time-reversal breaking parameter $\chi_{t}$ is increased, the two peaks that for $\chi_{t} = 0$ appear at $\Delta_{\pm}$ get closer together, merging into a single, smoother, peak at $eV = \Delta$ when $\chi_{t}$ reaches 1.
For $r = 0.5$ and $\chi_{t} = 0$ the conductance is maximized at $eV = \Delta_{\pm}$, but if $\chi_{t} = 1$ the peaks merge together into a single peak at $\Delta_{0}$. Notably, a small peak at zero bias remains even for the time-reversal symmetry breaking superconductors. This is in contrast with the zero bias conductance in junctions in which the time-reversal symmetry is broken by an exchange field, in which case the small peak disappears \cite{kokkeler2023anisotropic}. The difference can be explained as follows. Without an exchange field the correlations in the normal metal bar decay on a length scale $\sqrt{D/(2E)}$, irrespective of the phase difference between the s-wave and p-wave components of the pair potential, which are not mixed. On the other hand, in the presence of an exchange field, the singlet and triplet components are mixed, and the correlations decay on a length scale $\sqrt{D/(2(E\pm h))}$, where $h$ is the exchange field strength. Therefore, in the presence of an exchange,  crossed Andreev reflection is suppressed, but in SNN junctions with time-reversal symmetry breaking superconductors it is preserved.
%The difference between the two cases can be illustrated as follows. Without loss of generality, assume that $\vec{d}(\phi)$ points in the z-direction. The pair amplitudes $\langle\psi_{\uparrow}\psi_{\uparrow}\rangle$ and $\langle\psi_{\downarrow}\psi_{\downarrow}\rangle$ have a different phase if there is a phase difference between the singlet and triplet components. However, both phases are constant over space. Therefore the correlations with energy $E$ decay with characteristic length $\sqrt{D/(2E)}$, which diverges as $E\xrightarrow{}0$ whereas in the presence of an exchange field this phase difference varies with position and correlations decay with characteristic length $\sqrt{D/(2(E\pm h))}$, which does not diverge at $E = 0$. 
The results for junctions in which the p-wave component of the pair potential is dominant, e.g. $r = 2$ are shown in Fig. \ref{fig:TRBdeps}(b). The sharp change in $d\sigma/dV$ at $eV = |\Delta_{-}|$ and the conductance peak at $eV = \Delta_{+}$ likewise get closer together, merging into a small conductance dip at $eV = \Delta_{0}$ for $\chi_{t} = 1$. Moreover, an increase of $\chi_{t}$ decreases the zero bias conductance peak. 

\begin{figure}
    \centering
    \includegraphics[width = 8.6cm]{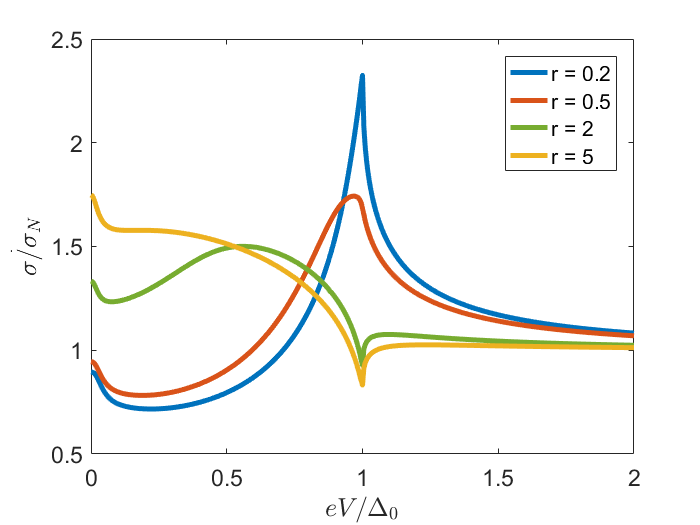}
    \caption{The conductance of the junction for different mixing parameters $r$ if $\chi_{t} = 1$. There is not a quantization of the zero bias conductance peak, unlike for $\chi_{t} = 0$, but rather a continuous change between a small zero bias conductance peak that does not exceed the normal state conductance and has a width on the order of the Thouless energy for $r<1$ and a larger peak for $r>1$ that exceeds the normal state conductance and may have a width larger than the Thouless energy. Other parameters are set to $\gamma_{B} = 2$, $z = 0.75$, $E_{\text{Th}}/\Delta_{0} = 0.02$.}\label{fig:RdepTRB1}
\end{figure}

\begin{figure*}
\centering
    {\hspace*{-2em}\includegraphics[width =8.6cm]{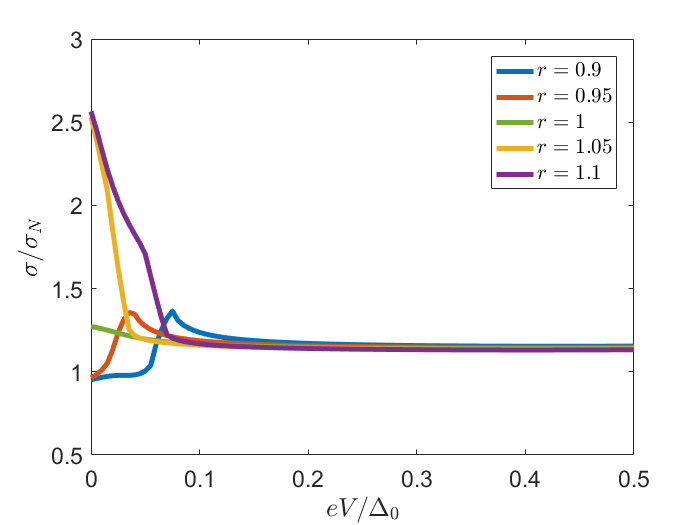}}
    \hfill
    {\hspace*{-2em}\includegraphics[width =8.6cm]{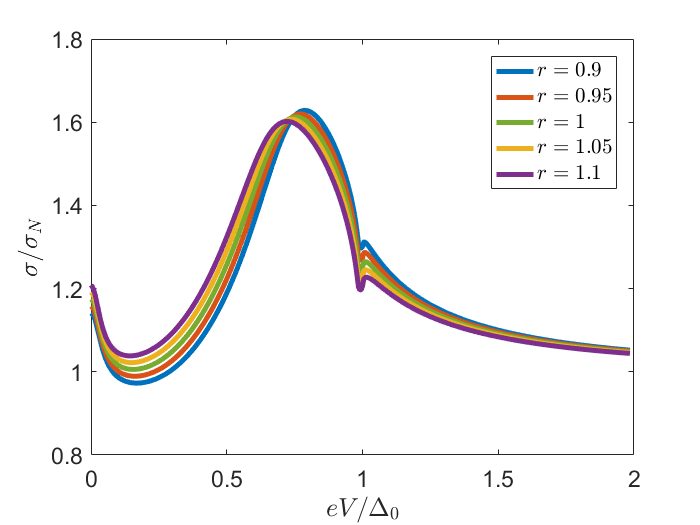}}
    \hfill
    {\hspace*{-1.5em}\includegraphics[width = 8.6cm]{figures/A.png}}
    \hfill
    {\hspace*{-2em}\includegraphics[width = 8.6cm]{figures/B.png}}
    \caption{The conductance near the transition between the s-wave dominated and p-wave dominated regime for s+helical (a) and is+helical (b) 
 p-wave superconductors. Other parameters are set to $\gamma_{B} = 2$, $z = 0.75$, $E_{\text{Th}}/\Delta_{0} = 0.02$.}\label{fig:TransitionHelical}
\end{figure*}

The dependence on the mixing parameter $r$ is illustrated in Figs. \ref{fig:RdepTRB1} and \ref{fig:TransitionHelical}. For an is+p-wave superconductor there is no quantization of the zero bias conductance, as shown in Fig. \ref{fig:RdepTRB1}. This can be understood as follows. The zero bias conductance depends on the quantities $\Delta_{\pm}/|\Delta_{\pm}|$. As discussed before, for $\chi_{t} = 1$ these quantities are complex, with the phase depending on the actual value of $r$, and therefore, the zero bias conductance depends non-trivially on $r$. This is in contrast with the case $\chi_{t} = 0$, when $\Delta_{\pm}/|\Delta_{\pm}|$ are both real, giving the quantization of conductance \cite{ikegaya2016quantization,kokkeler2023anisotropic}.
An analytical description  can be found in the 1D case. 
The matrix  $\check{C}(E =0,r)$ entering the  boundary condition Eq. \ref{eq:Cdef} is for is+helical p-wave superconductors given by
\begin{align}
    \check{C}(E = 0,r)&= \begin{bmatrix}-ir\sigma_{x} &\sqrt{r^2+1}\mathbf{1}_{\sigma}\\\sqrt{r^2+1}\mathbf{1}_{\sigma}&ir\sigma_{x}\end{bmatrix},
\end{align}
where $\mathbf{1}_{\sigma}$ is the identity matrix in spin space and $\sigma_{x}$ is the first Pauli matrix in spin space.
For finite $r$ the quantity $C$ is finite, showing that there is no pole at $E = 0$, in contrast to cases of a one-dimensional p-wave superconductor \cite{tanaka2022theory} or a two-dimensional s+helical p-wave superconductor \cite{kokkeler2023spin}. This means there is no zero energy Andreev bound state (ZESABS) for is+p-wave superconductors. The boundary quantity $C$ changes continuously as a function of $r$, and  in the limit $r\xrightarrow{}\infty$ the p-wave expression is recovered, in the limit $r\xrightarrow{} 0$ the s-wave expression is recovered, showing that the zero bias conductance continuously changes from the value attained by s-wave superconductors to the value attained by p-wave superconductors.

The broadness of the peak can be understood by considering the poles of $C$, which is connected to the existence of Andreev bound state in the clean limit \cite{tanaka2022theory}. The quantity $C$ is for is+p-wave superconductors given by
\begin{align}
&C(\phi) = H_{+}^{-}-H_{+}^{-1}H_{-} = \frac{1}{E^{2}-(\Delta_{s}^{2}+\Delta_{p}^{2}\sin^{2}\phi)}\nonumber\\&\Bigg(\sqrt{E^{2}-\Delta_{0}^{2}}\begin{bmatrix}
        E&\Delta_{isp+}\\\Delta_{isp-}&-E
    \end{bmatrix}+\nonumber\\&\begin{bmatrix}
        \Delta_{isp+}\Delta_{p}\cos\phi\sigma_{x}&-E\Delta_{p}\cos\phi\sigma_{x}\\-E\Delta_{p}\cos\phi\sigma_{x}&-\Delta_{isp+}\Delta_{p}\cos\phi\sigma_{x}
    \end{bmatrix}\Bigg),\nonumber\\
    &\Delta_{isp\pm} = i\Delta_{s}+\Delta_{p}\sin\phi\sigma_{y}
\end{align}
From this we infer that $C(\phi)$ has poles at $E = \Delta_{0}\sqrt{\frac{1+r^{2}\sin^{2}\phi}{1+r^{2}}}$. Therefore, the conductance peak is enhanced between $\Delta_{s} = \frac{\Delta_{0}}{\sqrt{r^{2}+1}}<eV<\Delta_{0}$. Thus, with increasing mixing parameter $r$ the peak width increases, as confirmed by the results in Fig. \ref{fig:RdepTRB1}. 

The continuity of $C$ as a function of $r$ implies that the transition between the s-wave dominated as p-wave dominated regimes is vastly different for the s+helical and is+helical p-wave superconductors, as shown in Fig. \ref{fig:TransitionHelical}.
For the s+helical p-wave superconductors there is a  topological phase transition at $r = 1$ \cite{tanaka2009theory}, which leads to a discontinuity of the conductance at zero bias \cite{kokkeler2023spin}, as shown in Fig. \ref{fig:TransitionHelical}(a). On the other hand, for is+helical p-wave superconductors the topological protection is absent and the conductance is continuous as a function of $r$ at $r =1$, see Fig. \ref{fig:TransitionHelical}(b). The conductance for $r\approx 1$ has similarities with both the conductance in s-wave junctions and the conductance in p-wave junctions. The zero bias conductance for $r\approx 1$ is larger than the normal state conductance, as for p-wave junctions. However, the peak at $eV = \Delta_{0}$ is distinctly larger than the one at $eV = 0$, which is reminiscent of s-wave junctions.
\section{Chiral superconductors}
The calculations were repeated for chiral superconductors. Chiral superconductors are distinctly different from helical superconductors, time reversal symmetry is broken for the p-wave state, even in the absence of an s-wave component of the pair potential. In chiral superconductors, the phase of the p-wave correlations depends on the angle $\phi$ with the normal of the surface, the d-vector is given by
\begin{align}
    \vec{d}(\phi) = e^{i\phi}(0,0,1).
\end{align}
The phase difference between the s-wave and the p-wave pair amplitudes will be defined as the phase difference for the singlet and triplet correlations for the mode normal to the interface.
For all calculations we used the same set of parameters as for the helical p-wave superconductors and B-W superconductors, that is, $\gamma_{B} = 2$, $z = 0.75$ and $\Delta_{0}/E_{\text{Th}} = 50$. 

%An important note with respect to s+chiral superconductors is that, contrary to the previous cases the phase difference between the s-wave and the p-wave components is not the same for all modes since the phase of the p-wave component is related to the direction of momentum. Therefore, considering a bulk material an s+chiral superconductor and an is+chiral superconductor are in fact the same. However, in an SN junction one can still define the phase $\frac{\pi}{2}\chi_{t}$ as the phase difference between the s-wave and p-wave component for the mode normal to the surface between the two materials. S
For (i)s+chiral p-wave superconductors we define $\chi_{t}$ using the phase difference of the mode normal to the interface. The terms s+chiral p-wave or is+chiral p-wave superconductors will refer $\chi_{t} = 0$ and $\chi_{t} = 1$. In practice this means that a single material can act both as an s+chiral and an is+chiral superconductor, for example in a setup with two disjointed SN junctions that make an angle of $\frac{\pi}{2}$ \cite{kokkeler2023anisotropic}. 
% \TK{Like in the setup we proposed in a previous paper}

\subsection{Density of states and pair amplitudes}
\begin{figure*}
    \centering
    {\hspace*{-2em}\includegraphics[width = 8.6cm]{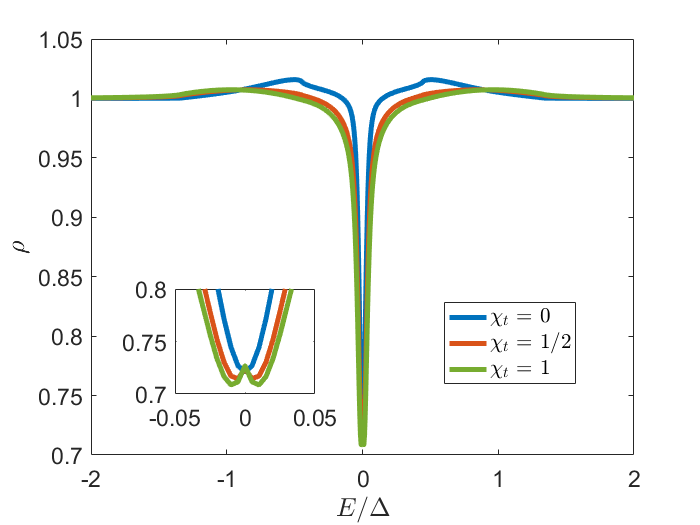}}
    {\hspace*{-2em}\includegraphics[width = 8.6cm]{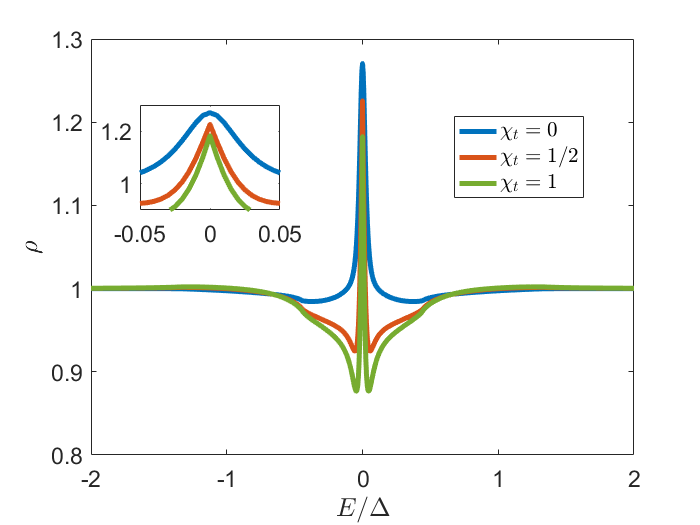}}
    {\hspace*{-1.5em}\includegraphics[width = 8.6cm]{figures/A.png}}
    {\hspace*{-2em}\includegraphics[width = 8.6cm]{figures/B.png}}
    \caption{The density of states in the (i)s+chiral p-wave superconductor junctions for different $\chi_{t}$. If $r = 0.5$ (a) there is a zero energy dip, if $r = 2$ (b) there is a zero energy peak. Other parameters are set to $\gamma_{B} = 2$, $z = 0.75$, $E_{\text{Th}}/\Delta_{0} = 0.02$.}\label{fig:LDOSchiral}
\end{figure*}

The local density of states for s+chiral p-wave junctions is shown in Fig. \ref{fig:LDOSchiral}. The results are similar to those of s+helical p-wave junctions, although the effects of $\chi_{t}$ are smaller for (i)s+chiral p-wave junctions. If the s-wave component of the pair potential is dominant the zero energy density of states is well below the normal density states, as shown in Fig. \ref{fig:LDOSchiral}(a). If the p-wave component of the pair potential is dominant, see Fig. \ref{fig:LDOSchiral}(b), the structure with both a broad dip and a narrow peak appears when $\chi_{t}$ is nonzero. It differs from the s+helical p-wave case by having a distinctly broader dip and a zero energy density of states that is well above the normal density of states.

\begin{figure*}
    \centering
    {\hspace*{-2em}\includegraphics[width = 4.3cm]{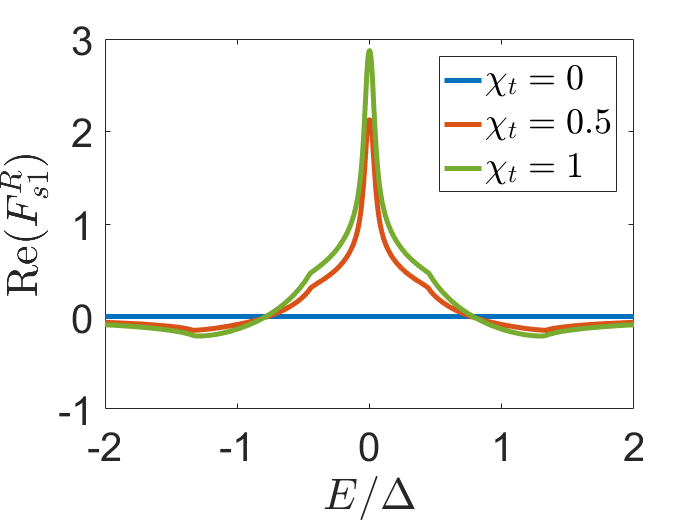}}
    \hfill
    {\hspace*{-2em}\includegraphics[width = 4.3cm]{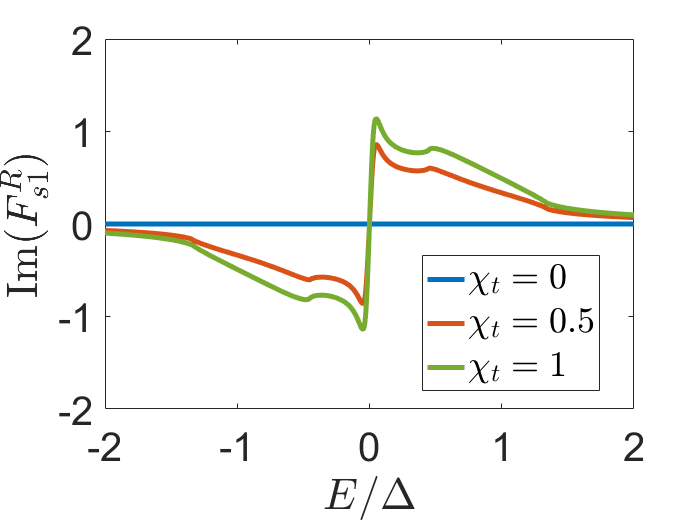}}
    \hfill
    {\hspace*{-2em}\includegraphics[width = 4.3cm]{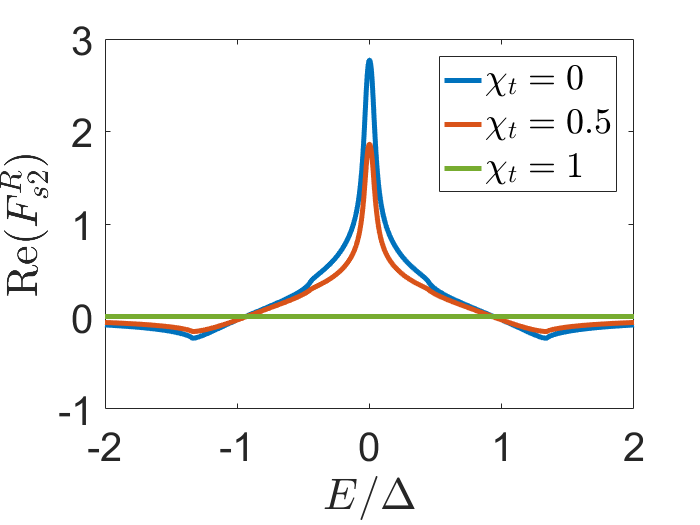}}
    \hfill
    {\hspace*{-2em}\includegraphics[width = 4.3cm]{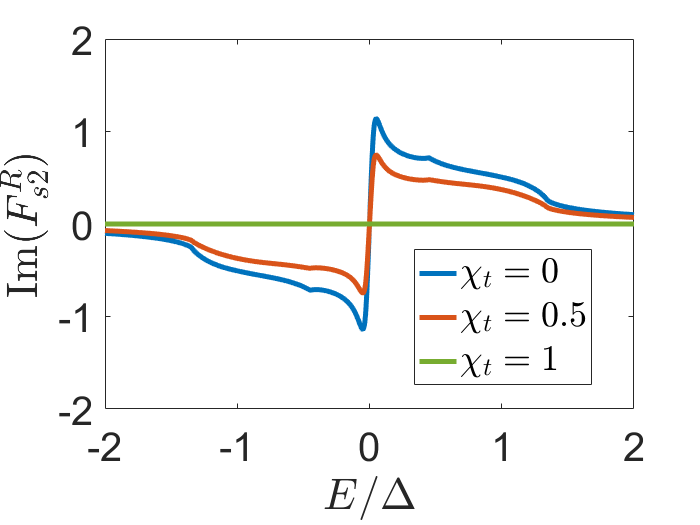}}
    \hfill
    {\hspace*{-2em}\includegraphics[width = 4.3cm]{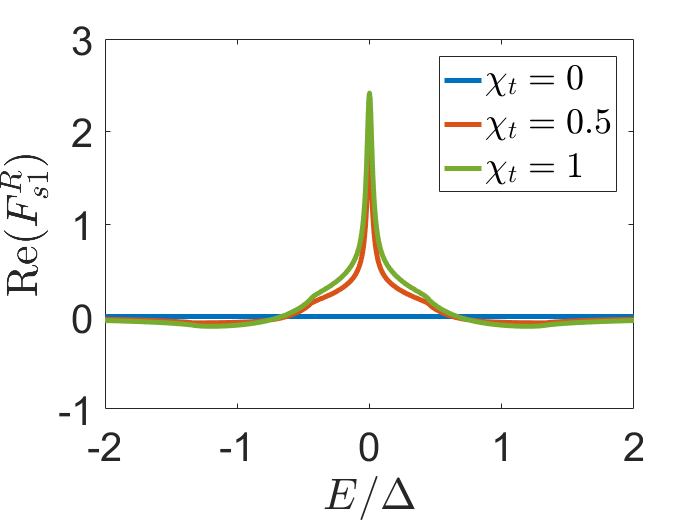}}
    \hfill
    {\hspace*{-2em}\includegraphics[width = 4.3cm]{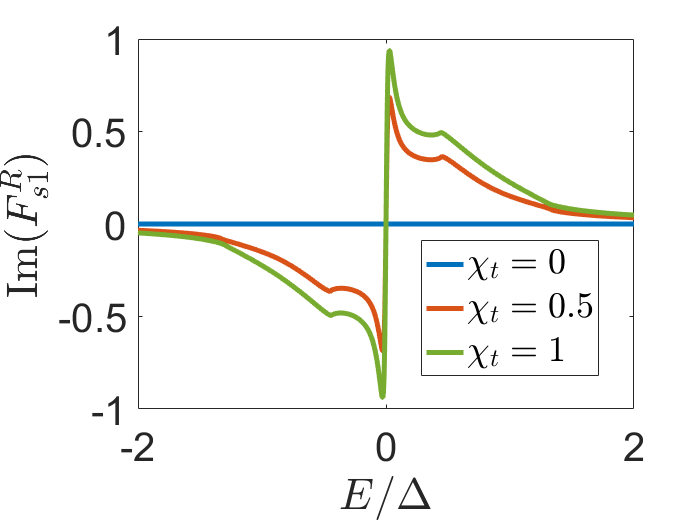}}
    \hfill
    {\hspace*{-2em}\includegraphics[width = 4.3cm]{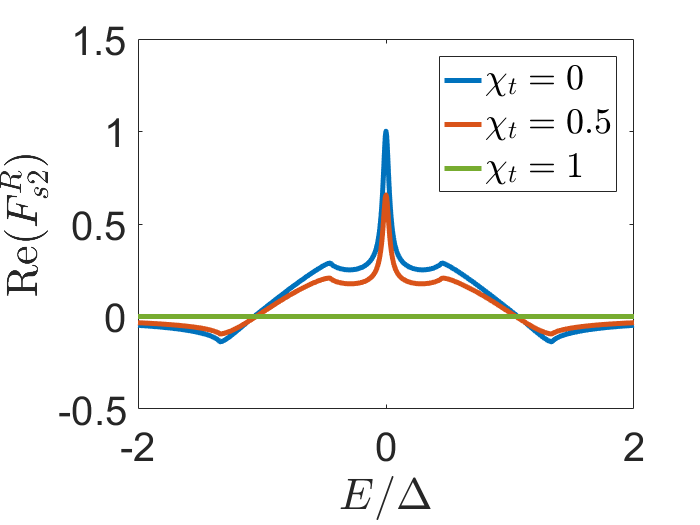}}
    \hfill
    {\hspace*{-2em}\includegraphics[width = 4.3cm]{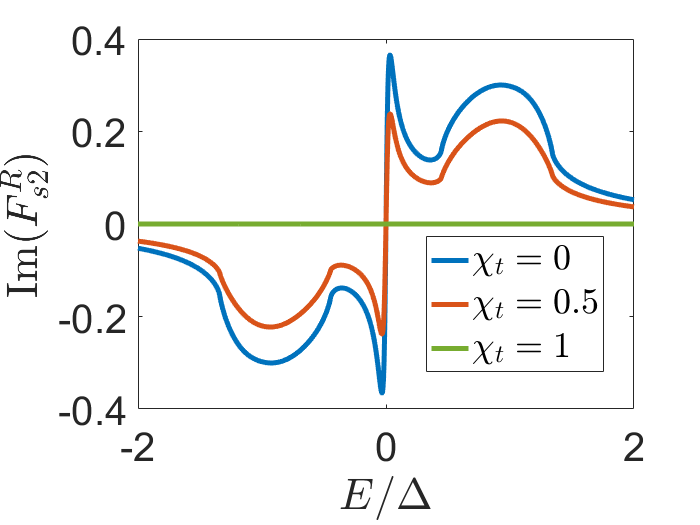}}
    \caption{The singlet pair amplitudes for $e^{i\chi_{t}\frac{\pi}{2}}s$+chiral p-wave superconductors for s-wave dominant superconductors with $r = 0.5$ (top row) and p-wave dominant superconductors with $r = 2$ (bottom row). Other parameters are set to $\gamma_{B} = 2$, $z = 0.75$, $E_{\text{Th}}/\Delta_{0} = 0.02$. }\label{fig:ChiralSingletPA}
\end{figure*}

\begin{figure*}
    \centering
    {\hspace*{-2em}\includegraphics[width = 4.3cm]{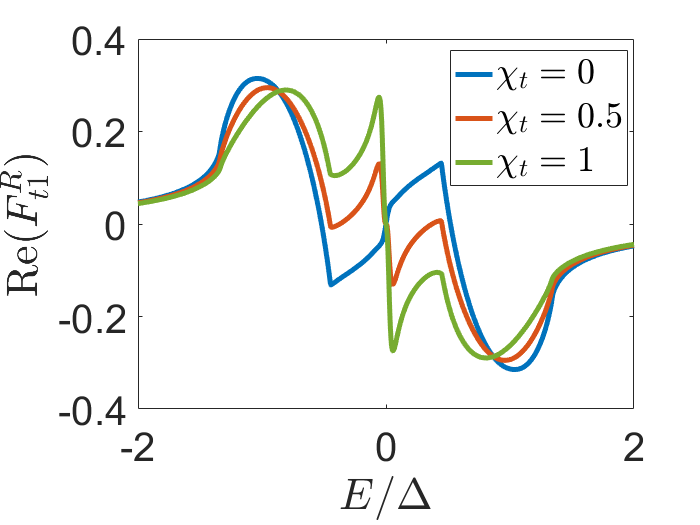}}
    \hfill
    {\hspace*{-2em}\includegraphics[width = 4.3cm]{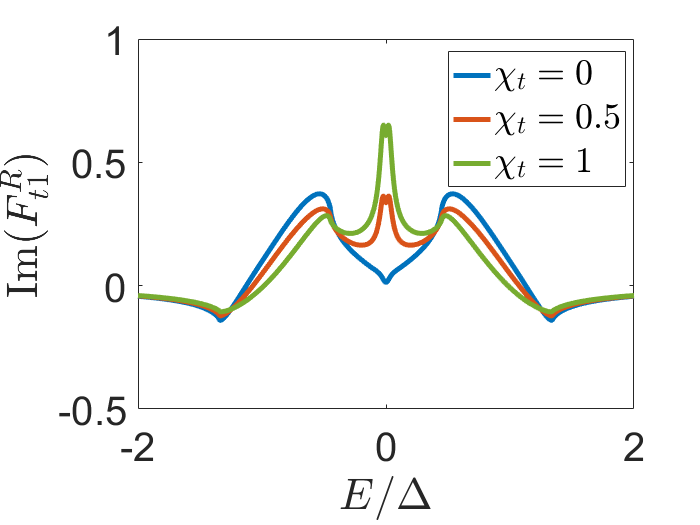}}
    \hfill
    {\hspace*{-2em}\includegraphics[width = 4.3cm]{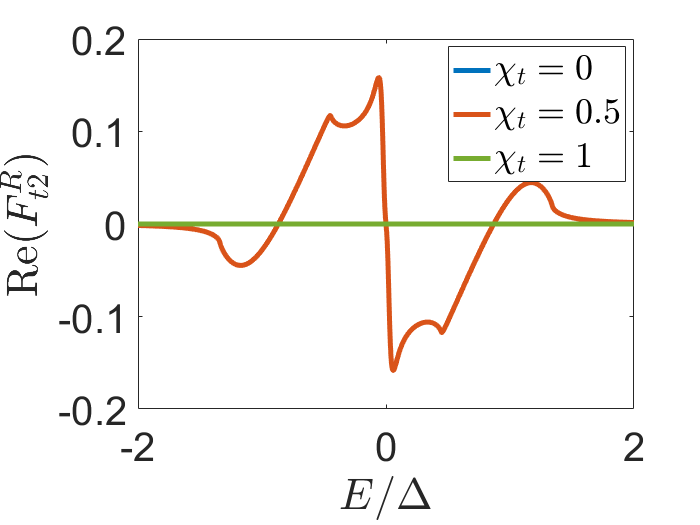}}
    \hfill
    {\hspace*{-2em}\includegraphics[width = 4.3cm]{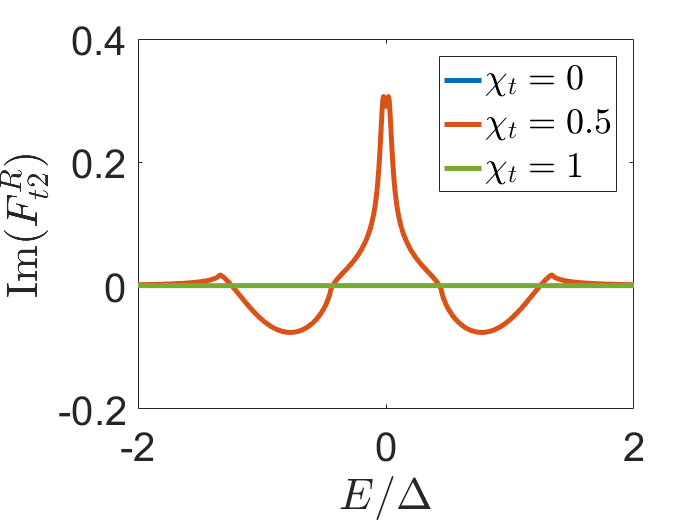}}
    \hfill
    {\hspace*{-2em}\includegraphics[width = 4.3cm]{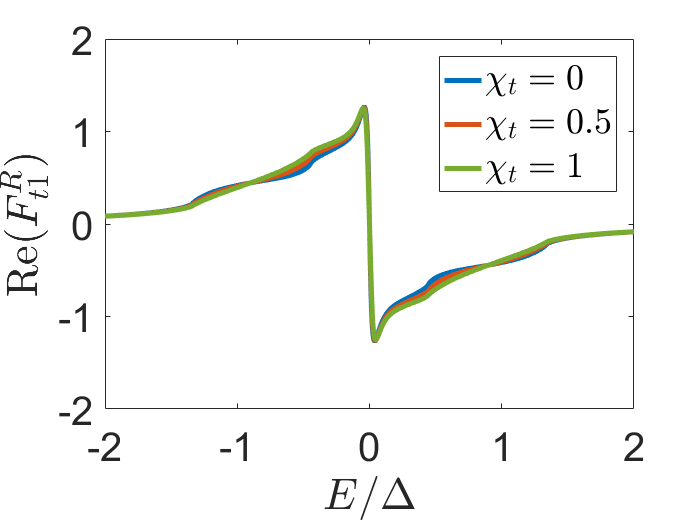}}
    \hfill
    {\hspace*{-2em}\includegraphics[width = 4.3cm]{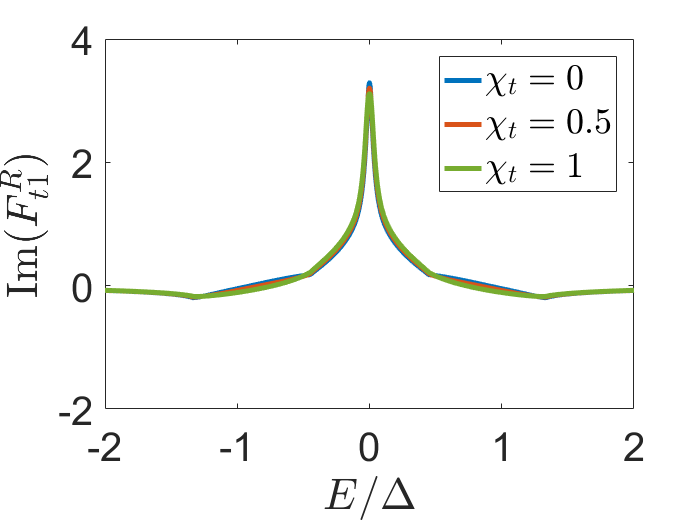}}
    \hfill
    {\hspace*{-2em}\includegraphics[width = 4.3cm]{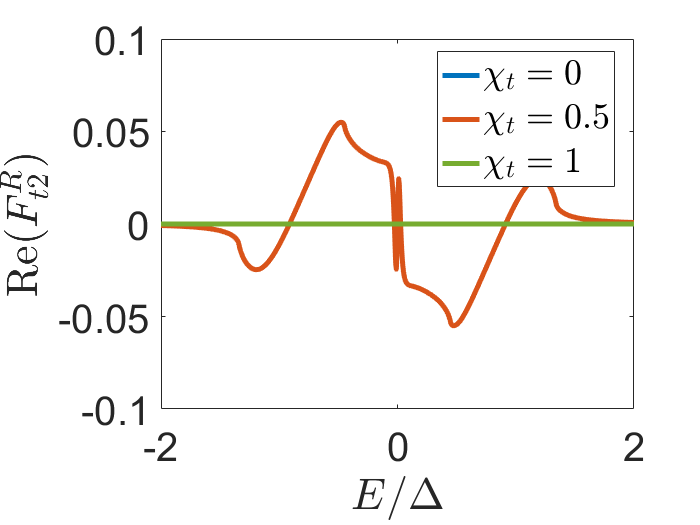}}
    \hfill
    {\hspace*{-2em}\includegraphics[width = 4.3cm]{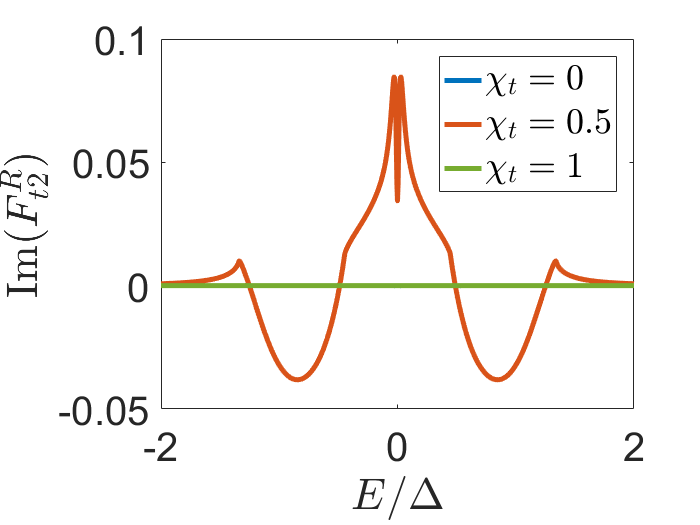}}
    \caption{The triplet pair amplitudes for $e^{i\chi_{t}\frac{\pi}{2}}s$+chiral p-wave superconductors for s-wave dominant superconductors with $r = 0.5$ (top row) and p-wave dominant superconductors with $r = 2$ (bottom row). Other parameters are set to $\gamma_{B} = 2$, $z = 0.75$, $E_{\text{Th}}/\Delta_{0} = 0.02$. }\label{fig:ChiralTripletPA}
\end{figure*}

The pair amplitudes are shown in Figs. \ref{fig:ChiralSingletPA} and \ref{fig:ChiralTripletPA}. The magnitudes of the dominant pair amplitude (Figs. \ref{fig:ChiralSingletPA}(top) and \ref{fig:ChiralTripletPA}(bottom)) are, similarly to the helical case, very similar for different $\chi_{t}$, though the phase of the induced correlations strongly depends on $\chi_{t}$.
On the other hand, the results for the subdominant pair amplitude, depend strongly on $\chi_{t}$ and they are distinctly different from the s+helical p-wave superconductors. First of all, even if $\chi_{t} = 0$ the subdominant component need not vanish at zero energy. Indeed, $\text{Re}(F_{s2}(E = 0))\neq 0$ for triplet dominant pair potentials, see Figs. \ref{fig:ChiralSingletPA}(bottom), and similarly $\text{Im}(F_{t1,2})(E = 0)\neq 0$ for singlet dominant junctions, as shown in Fig. \ref{fig:ChiralTripletPA}(top).
This is due to the time-reversal symmetry breaking of the chiral p-wave component of the pair potential. Next to this, even for $\chi_{t} = 1$ there is a small and narrow dip around zero energy in the triplet pair amplitude.
The singlet pair amplitude has a zero energy peak even for triplet dominant s+p-wave superconductors. This peak increases when $\chi_{t}$ is increased.

These effects can be attributed to the angle dependence of the phase difference between the s-wave and p-wave components of the pair potential. Whereas for the helical p-wave junction this phase difference is $\chi_{t}$ for all modes, for the chiral p-wave case the phase difference between singlet and triplet components varies between $\chi_{t}-\frac{\pi}{2}$ and $\chi_{t}+\frac{\pi}{2}$. Therefore the results are a weighted average over this range of phase differences.
Because the transmission for normal incidence is larger than for oblique incidences there is a difference between s+chiral p-wave and is+chiral p-wave superconductors, but it is less pronounced than for (i)s+helical p-wave superconductors. All pair amplitudes can exist at zero energy for any phase difference between the s-wave and p-wave components of the pair potential.
\subsection{Conductance}
\begin{figure*}
    \centering
    {\hspace*{-1.5em}\includegraphics[width = 8.6cm]{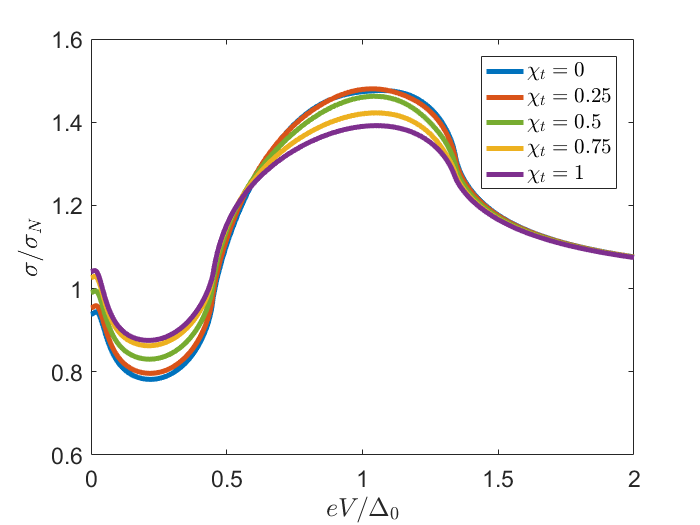}}
    \hfill
    {\hspace*{-2em}\includegraphics[width = 8.6cm]{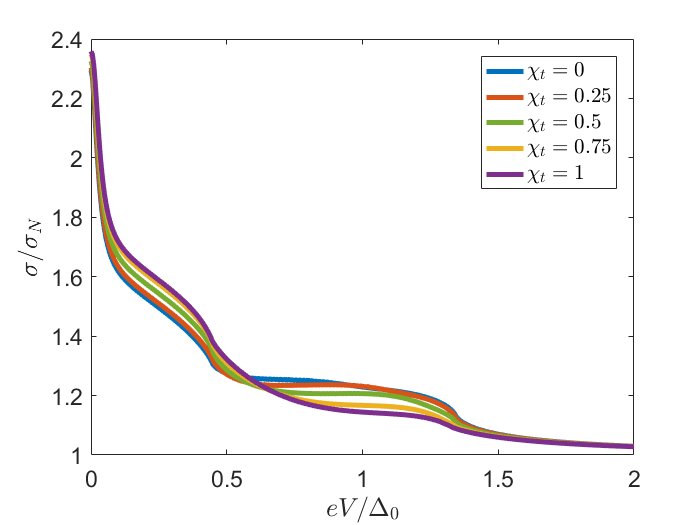}}
    \hfill
    \includegraphics[width = 8.6cm]{figures/A.png}
    \hfill
    \includegraphics[width = 8.6cm]{figures/B.png}
    \caption{The dependence of the conductance on $\chi_{t}$ for s-wave dominant (left, $r = 0.5$) and p-wave dominant (right, $r = 2$) superconductors. Other parameters are set to $\gamma_{B} = 2$, $z = 0.75$, $E_{\text{Th}}/\Delta_{0} = 0.02$.}\label{fig:chitdep}
\end{figure*}

The conductance in s+chiral p-wave superconductor junctions is different compared to s+helical p-wave junctions.
First of all, for the s+chiral p-wave superconductor there are no sharp peaks at $eV = \Delta_{\pm}$ because the eigenvalues of the matrix pair potential are complex and angle dependent even for $\chi_{t} = 0$ \cite{burset2014transport}. Thus, there is no transition between two sharp peaks and a single broad one when using (i)s+chiral p-wave superconductors. Instead, as shown in Fig. \ref{fig:chitdep}, the zero bias conductance is slightly increased by an increase of $\chi_{t}$, whereas the conductance for $|\Delta_{-}|<eV<\Delta_{+}$ is slightly decreased with increasing $\chi_{t}$.

\begin{figure*}
    \centering
    {\hspace*{-1.5em}\includegraphics[width = 8.6cm]{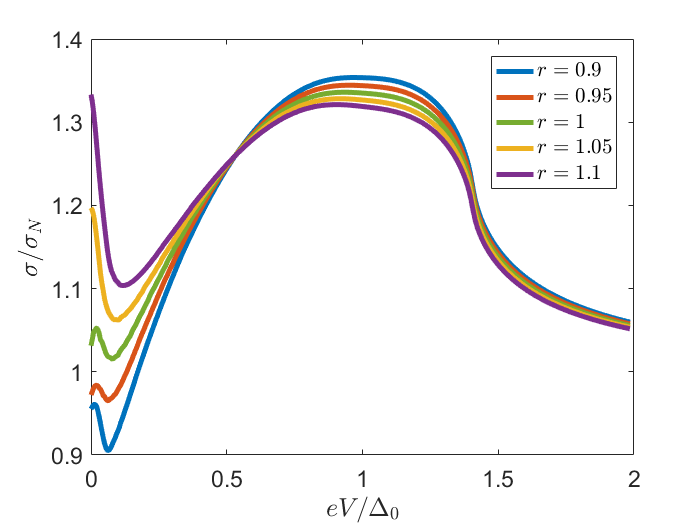}}
    \hfill
    {\hspace*{-2em}\includegraphics[width = 8.6cm]{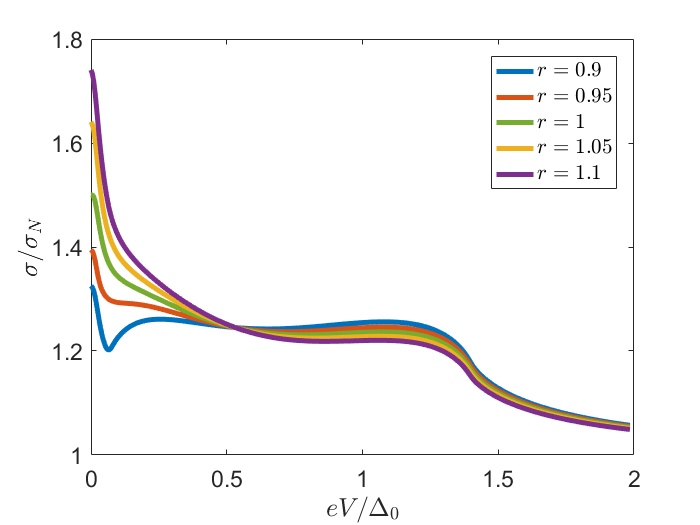}}
    \hfill
    \includegraphics[width = 8.6cm]{figures/A.png}
    \hfill
    \includegraphics[width = 8.6cm]{figures/B.png}
    \caption{The transition between the regimes of s-wave dominant and p-wave dominant superconductors for the s+chiral p-wave (a) and is+chiral p-wave superconductors (b). Other parameters are set to $\gamma_{B} = 2$, $z = 0.75$, $E_{\text{Th}}/\Delta_{0} = 0.02$.}\label{fig:TransitionChiral}
\end{figure*}

The difference between the s+chiral p and is+chiral p-wave superconductors is most clearly visible around $r = 1$, the transition point between s-wave dominant and p-wave dominant superconductors. In contrast with s+helical p-wave superconductor junctions, there is no discontinuity in the conductance in either case. For the s+chiral p-wave superconductor ($\chi_{t} = 0$), as shown in Fig. \ref{fig:TransitionChiral}(a), the conductance is more similar to the conductance of a junction with an s-wave superconductor, with the zero bias conductance rapidly increasing as a function of $r$, and the broad peak around $eV = \Delta_{0}$ decreasing slowly with $r$. On the other hand, for the is+chiral p-wave superconductor ($\chi_{t} = 1$), as shown in Fig. \ref{fig:TransitionChiral}(b), the conductance is more similar to a p-wave dominated superconductor. Also for this case, the zero bias conductance highly depends on $r$, whereas the peak around $eV = \Delta_{0}$ develops slowly with decreasing $r$.  
Summarizing, for the (i)s+chiral p-wave superconductors, the influence of the phase difference $\chi_{t}$ between the s-wave and p-wave components of the pair potential on all properties of the junction is smaller because the phase difference between the singlet and triplet correlations is different for each mode that contributes.
\section{3d B-W superconductors}
Time-reversal and inversion symmetry broken superconductors  are not restricted to two-dimensional materials, but may also appear in three dimensions \cite{kanasugi2022anapole}.
Therefore we generalized the formalism to include superconductors whose order parameter depends on $k_{x,y,z}$.  %Second, we study the singlet chiral d-wave superconductor with $\Delta = \Delta_{0}e^{i(k_{x}+ik_{y})}\sin{k_{z}}$. 
For the 3D superconductor, we need to extend the Tanaka-Nazarov boundary conditions. This extension is  straightforward. The reflected mode has opposite $k_{y}$ and $k_{z}$ instead of only opposite $k_{y}$, and integration needs to be performed over a half-sphere instead of a half-circle. It is assumed that the transmission only depends on the angle made with the surface normal. The inclusion of 3D angles in the boundary condition does make the code more computationally expensive. However, for certain types of three-dimensional p-wave superconductors symmetry can be exploited to reduce the computational costs.

Specifically, we consider the proximity effect induced by a superconductor with a B-W pair amplitude. The proximity effect of B-W superconductors was first studied for the chargeless Helium superfluid \cite{higashitani2009proximity}. For the B-W phase the d-vector is given by $\vec{d}(\phi,\psi) = (\cos\psi,\sin\psi\sin\phi,\sin\psi\cos\phi)$. The B-W phase is therefore the natural extension of the 2D helical superconductor to a 3D material. The magnitude of the gap is constant over the Fermi surface, and the direction of the d-vector is the same as the direction of momentum. The results are therefore predicted to be similar as well.

%\subsection{B-W phase}

For the 2D helical superconductors it has been shown following symmetry arguments for the modes at $(k_{x},\pm k_{y})$ in the absence of a magnetic field only singlet correlations and triplet correlations with d-vector proportional to $\langle d\rangle\cdot\sigma = \sigma_{x}$ are induced \cite{kokkeler2023spin}. Because in 3D B-W superconductors $k_{y}$ and $k_{z}$ are equivalent symmetry dictates that this feature holds for 3D B-W superconductors as well. Even more, this equivalence dictates that the singlet and $\sigma_{x}$-triplet contributions of all modes with fixed $k_{x}$ are the same. The $\phi$-integral can thus be performed analytically and yields a factor $2\pi$.  Therefore, for the B-W superconductor we do not need to numerically integrate over two dimensions and the numerical costs are comparable to those for junctions with a 2D superconductor.
To improve convergence, a Dynes parameter $\Gamma = 0.01\Delta_{0}$ was added as an imaginary part of the energy.

%that for any $\psi$ the sum of the contributions at $(\cos\phi,\sin\phi\sin\psi,\sin\phi\cos\phi)$ and at $(\cos\phi,-\sin\phi\sin\psi,-\sin\phi\cos\phi)$ only has $\sigma_{x}$ and $\mathbf{1}$ contributions, and these contributions are in fact independent of $\psi$. For this case therefore only a sum over the line with $k_{z} = 0$ is needed. 
%Moreover, the The difference with the 2D case is thus only that each contribution should be multiplied by a factor $2\sin\phi$, where the $\sin\phi$ is proportional to the radius of the circle $k_{y}^{2}+k_{z}^{2} = \sin^{2}\phi$ and the factor 2 appears because the surface of a unit sphere is $4\pi$ whereas the circumference of the circle is $2\pi$.

\begin{figure}
\centering
    \includegraphics[width = 8.6cm]{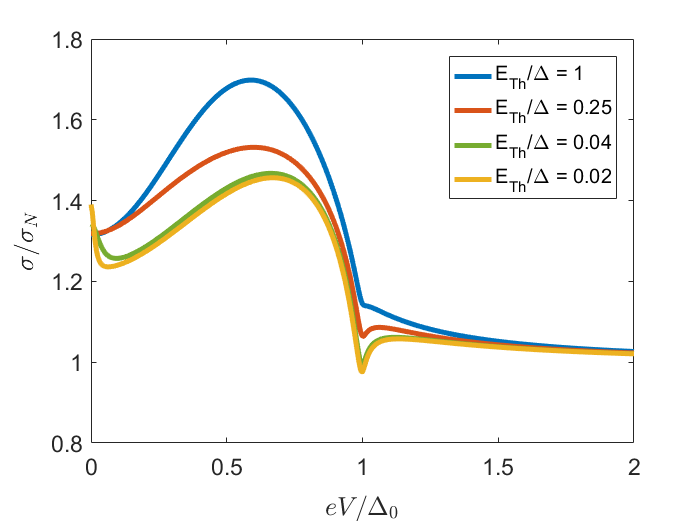}
    
    \caption{The conductance in an SNN junction with a B-W phase p-wave superconductor for different lengths of the junction with $z = 0.75$ and $\gamma_{B} = 2$. For longer junctions, the peak at zero energy is larger and sharper. For larger voltages $\sigma/\sigma_{N}$ decreases with increasing length. }\label{fig:BWp}
\end{figure}

First, we consider the case of a p-wave B-W superconductor, that is, without inclusion of any s-wave component in the pair potential. The conductance is shown for different lengths in Fig. \ref{fig:BWp}. In long junctions, there is a sharp zero bias conductance peak with a width on the order of the Thouless energy. For short junctions, however, there exists no zero bias conductance peak. This is in contrast with other types of p-wave superconductors such as the 2D helical and chiral superconductors described in previous sections. 
%Even though the zero bias conductance peak is absence, the value of the zero bias conductance is larger than the normal state conductance. This can still be used to detect the presence of unconventional superconductivity.

\begin{figure}
    \centering
    \includegraphics[width= 5.6cm]{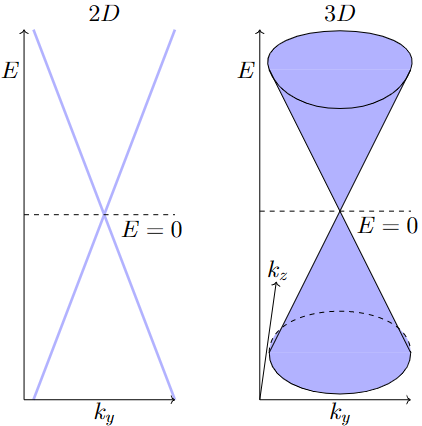}
    \caption{The dispersion of the Andreev bound states for 2D helical and 3D B-W p-wave superconductors. In both cases, the energy of the bound states is linear in the magnitude of the momentum. For the 2D case, this remains that the density of Andreev bound states is constant around $E = 0$, for the 3D case the density of Andreev bound states vanishes at $E = 0$. Parameters are set to $\gamma_{B} = 2$, $z = 0.75$, $E_{\text{Th}}/\Delta_{0} = 0.02$.}
    \label{fig:ABS2vs3}
\end{figure}

The absence of a zero bias conductance peak  in the short limit can be understood by considering the tunneling via Andreev bound states in ballistic junctions, which has a much wider range of possibilities for three-dimensional superconductors compared to their two-dimensional counterparts \cite{buchholtz1981identification,asano2003a,lofwander2001andreev,yamakage2012theory,tamura2017theory,fu2010odd}. The dispersion of the Andreev bound states in junctions with helical or B-W p-wave superconductors is shown in Fig. \ref{fig:ABS2vs3}.
In each case there is a zero energy Andreev state at normal incidence, and the energy is approximately linear in the magnitude of momentum. For 2D superconductors this means that the density of Andreev bound states is approximately constant and finite around $E = 0$ \cite{matsumoto1999quasiparticle,furusaki2001spontaneous}, leading to a broad conductance peak centered at $eV = 0$. On the other hand, for 3D superconductors, the density of Andreev bound states vanishes at $E = 0$ and is finite for nonzero energies. This leads to an enhancement of the conductance for nonzero voltages compared to the zero bias conductance. The balance between this effect and coherent Andreev reflection determines whether there is a zero bias conductance peak or zero bias conductance dip in junctions with B-W superconductors. %For 3D superconductors the mode at normal incidence has a much smaller effect than for 2D superconductors because the phase space is much larger. Therefore the contribution of modes at oblique incidence, which have Andreev bound states at nonzero energies, becomes more pronounced. This leads to the absence of a zero energy peak.


\begin{figure*}
    \centering
    {\hspace*{-2em}\includegraphics[width = 8.6cm]{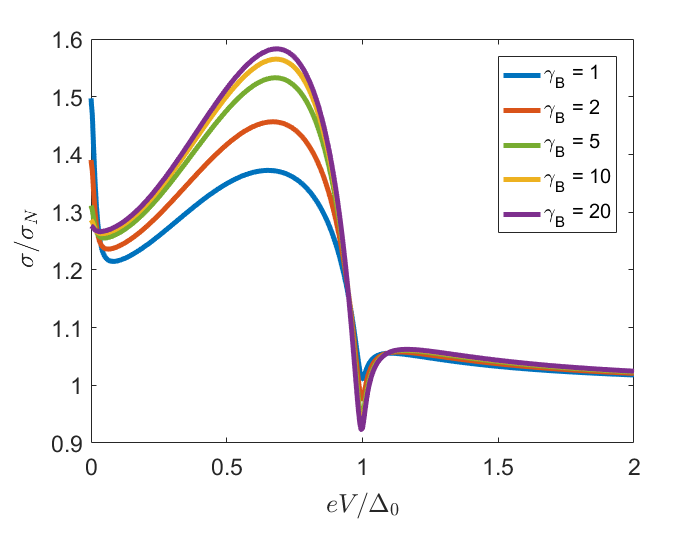}}
    \hfill
    {\hspace*{-2em}\includegraphics[width = 8.6cm]{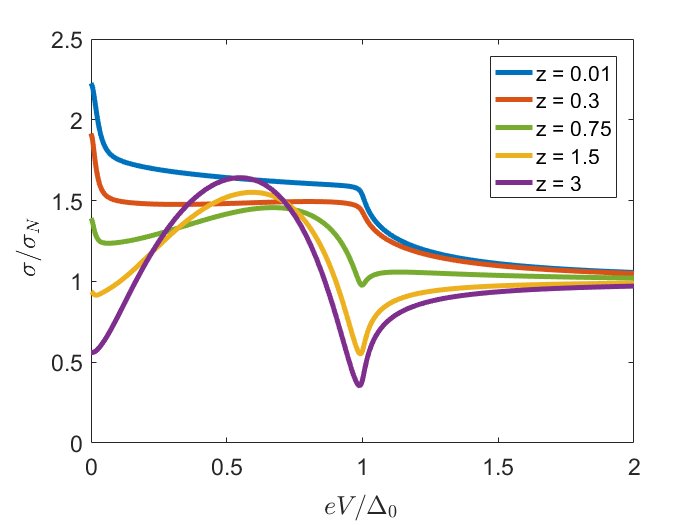}}
    \hfill
    {\hspace*{-1.5em}\includegraphics[width = 8.6cm]{figures/A.png}}
    \hfill
    {\hspace*{-2em}\includegraphics[width = 8.6cm]{figures/B.png}}
    \caption{The dependence of the conductance on the boundary parameters $\gamma_{B}$ (left) and $z$ (right). The Thouless energy was set to $E_{\text{Th}}/\Delta_{0} = 0.02$ in all calculations. In the left figure $z = 0.75$, in the right figure $\gamma_{B} = 2$. An increase of $\gamma_{B}$ leads to a decrease of the zero bias conductance, but an increase of the conductance peak between $eV = 0.5\Delta_{0}$ and $eV = \Delta_{0}$. For small $z$ the conductance decreases monotonically with voltage, whereas for large $z$ the peak between $eV = 0.5\Delta_{0}$ and $eV = \Delta_{0}$ becomes more pronounced, whereas there exists a zero bias conductance dip in this case.}\label{fig:gammadepBW}
\end{figure*}

As shown in Fig. \ref{fig:gammadepBW}(a) an increase of the boundary resistance, i.e. an increase $\gamma_{B}$ leads to a suppression of the zero bias conductance peak, but an increase in the conductance peak between $eV = 0.5\Delta_{0}$ and $eV = \Delta_{0}$, while the dip around $eV = \Delta_{0}$ is also more pronounced.
The conductance strongly depends on the $z$-parameter, as shown in Fig. \ref{fig:gammadepBW}(b). For small $z$, that is, an interface with low transparency, the zero bias conductance consists of the usual sharp peak and an almost flat peak, whereas for large $z$ a zero bias conductance dip is found. With increasing $z$ also the conductance for $eV\approx\Delta$ is strongly suppressed and may even decrease below the normal state conductance. These effects can be understood as follows. As $\gamma_{B}$ increases the boundary resistance increasingly dominates the total resistance. The coherent Andreev reflection suppresses the resistance inside the bar, but not the boundary resistance. Therefore, the coherent Andreev reflection peak is suppressed if the boundary resistance becomes more dominant.

\begin{figure*}
    \centering
    {\hspace*{-2em}\includegraphics[width = 8.6cm]{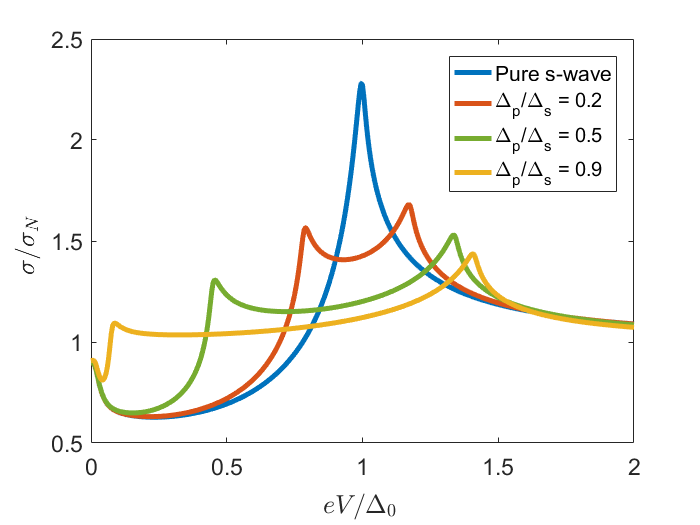}}
    \hfill
    {\hspace*{-2em}\includegraphics[width = 8.6cm]{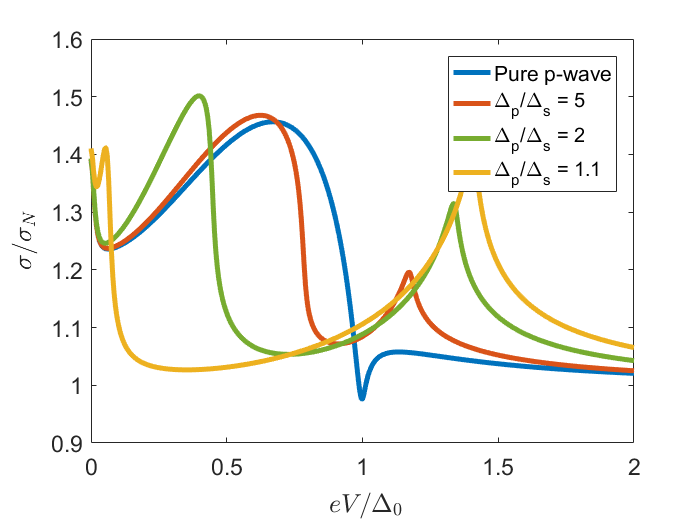}}
    \hfill
    {\hspace*{-1.5em}\includegraphics[width = 8.6cm]{figures/A.png}}
    \hfill
    {\hspace*{-2em}\includegraphics[width = 8.6cm]{figures/B.png}}
    \caption{The conductance of the s+B-W/N/N junction for different ratios of the s-wave and p-wave components of the pair potential. The zero bias conductance is quantized, it only depends on whether the s-wave or p-wave component of the pair potential is dominant. The height of the ZBCP is significantly suppressed compared to the two-dimensional s+helical p-wave junctions. The conductance has sharp features at $eV = E_{\text{Th}}$ and $eV = \Delta_{\pm}$. Parameters are set to $\gamma_{B} = 2$, $z = 0.75$, $E_{\text{Th}}/\Delta_{0} = 0.02$.}\label{fig:BWS}
\end{figure*}

\begin{figure*}
    \centering
    {\hspace*{-2em}\includegraphics[width = 8.6cm]{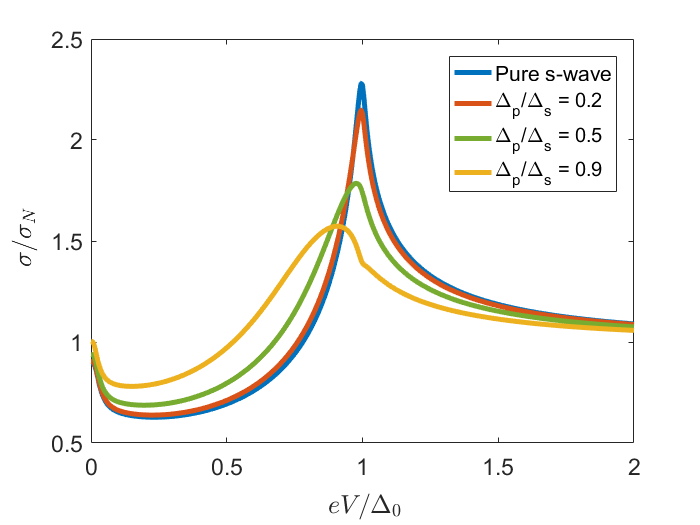}}
    \hfill
    {\hspace*{-2em}\includegraphics[width = 8.6cm]{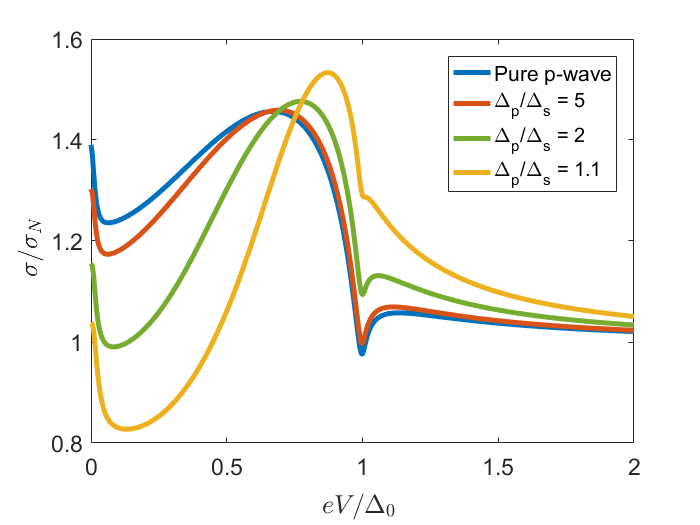}}
    \hfill
    {\hspace*{-1.5em}\includegraphics[width = 8.6cm]{figures/A.png}}
    \hfill
    {\hspace*{-2em}\includegraphics[width = 8.6cm]{figures/B.png}}
    \caption{The conductance of the is+B-W/N/N junction for different ratios of the s-wave and p-wave components of the pair potential. The zero bias conductance depends on the particular ratio of the s and p-wave components, crossing the normal state conductance at $r = 1$. Parameters are set to $\gamma_{B} = 2$, $z = 0.75$, $E_{\text{Th}}/\Delta_{0} = 0.02$.}\label{fig:BWIS}
\end{figure*}

Next, we considered the s+p-wave and is+p-wave junctions with $E_{\text{Th}}/\Delta_{0} =0.02$ and $z = 0.75$. The results in Fig. \ref{fig:BWS} for s+p-wave superconductors are shown and in Fig. \ref{fig:BWIS} for is+p-wave superconductors.
%Similar to the two-dimensional (i)s+p-wave pair potentials, if the s-wave component is dominant,
The main features are similar to that of the two-dimensional helical p-wave superconductors. As shown in Fig. \ref{fig:BWS}(a), if a subdominant B-W pair potential is added to an s-wave superconductor with the same phase, the peak at $eV = \Delta_{s}$ splits into two peaks at $eV = \Delta_{\pm}$. The zero bias conductance is not altered since the B-W superconductor, like the helical superconductor, is a topological superconductor protected by time-reversal symmetry \cite{tanaka2009theory}. On the other hand, if the subdominant B-W is added to an s-wave superconductor with a phase difference of $\pi/2$, a single peak remains, as highlighted in Fig. \ref{fig:BWIS}(a). This peak is suppressed, broadened and shifted towards slightly lower voltages. The zero bias conductance is almost constant but slightly increases since in the absence of time-reversal symmetry there is no topological protection.

The case in which the p-wave component of the pair potential is dominant is different compared to the 2D s+helical p-wave superconductors, since the conductance for a  3D p-wave superconductor is different from the conductance for a 2D p-wave superconductor as shown before in Fig. \ref{fig:BWp}. However, the main influence of the inclusion of an s-wave component in the pair potential at the same is similar, see Fig. \ref{fig:BWS}(b). The sharp feature in the conductance at $eV = \Delta_{p}$ splits into two peaks at $eV = \Delta_{p}\pm\Delta_{s}$, while the zero bias conductance is unaltered. For $\Delta_{p}-\Delta_{s}\ll\Delta_{p}$ this peak approaches the zero bias conductance peak. The two peaks are clearly distinguishable as long as $\Delta_{p}-\Delta_{s}\gg E_{\text{Th}}$. If time-reversal symmetry is broken, see Fig. \ref{fig:BWIS}(b), the zero bias conductance is decreased if an s-wave component is included in the pair potential, and the peak in the conductance at finite voltage becomes sharper and larger. Moreover, it increases towards $eV = \Delta_{0}$.
\section{Conclusions}
In conclusion,  we have shown that the phase difference that may appear between the singlet and triplet correlations in time-reversal symmetry-broken non-centrosymmetric superconductors has a large influence on the local density of states, pair amplitudes, and conductance in dirty SNN junctions. Novel features  are  particularly visible if the p-wave superconductor is of the 2D helical or 3D B-W type, topological superconductors protected by time-reversal symmetry. We have shown that in the absence of topological protection, the zero bias conductance varies continuously as a function of the mixing parameter between the s-wave and p-wave components of the pair potential. Our results provide a new way to detect time-reversal symmetry breaking in s+p-wave superconductors.
Next to this, they highlight that the breaking of topological protection significantly alters the proximity effect.

For (i)s+chiral p-wave superconductors the phase difference between singlet and triplet components is mode-dependent, and therefore the dependence of the density of states and conductance on the phase of the s-wave component of the pair potential is much weaker. Next to this, our results show that a zero bias conductance peak can be absent when using three-dimensional odd-parity superconductors. This clearly distinguishes three-dimensional odd-parity superconductors from their two-dimensional counterparts. 
\section{Acknowledgements}
T.K. and S.B.  acknowledge  financial support from Spanish MCIN/AEI/ 10.13039/501100011033 through project PID2020-114252GB-I00 (SPIRIT) and  TED2021-130292B-C42,
and the Basque Government through grant IT-1591-22.
\newpage
\bibliography{biblio}
\newpage
\appendix
\section{Density of states and pair amplitudes in the (i)s+B-W junction}
\begin{figure*}
    \centering
    {\hspace*{-2em}\includegraphics[width = 8.6cm]{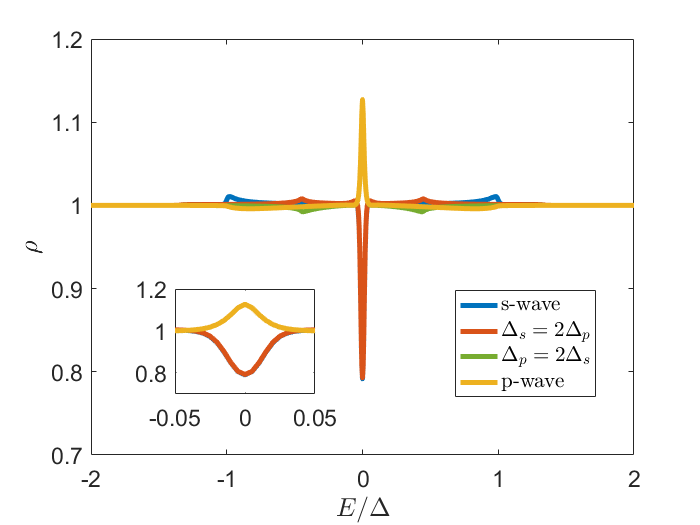}}
    \hfill
    {\hspace*{-2em}\includegraphics[width = 8.6cm]{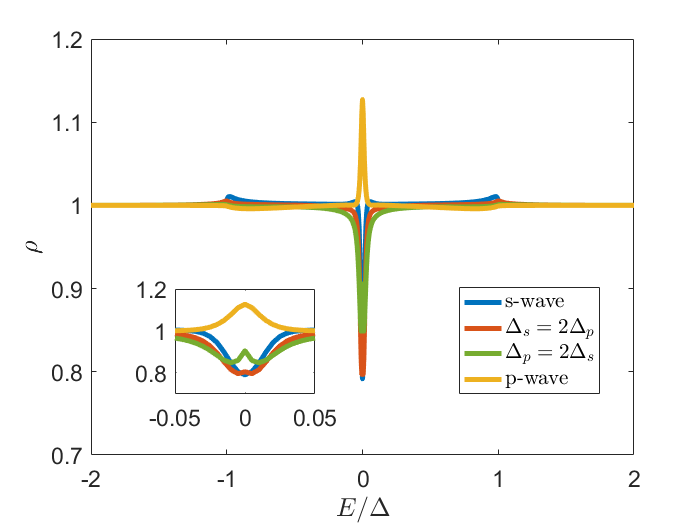}}
    \hfill
    {\hspace*{-1.5em}\includegraphics[width = 8.6cm]{figures/A.png}}
    \hfill
    {\hspace*{-2em}\includegraphics[width = 8.6cm]{figures/B.png}}
    \caption{The local density of states for s+p-wave (a) and is+p-wave (b) superconductors for s-wave dominant  and p-wave dominant  s+B-W superconductors.}\label{fig:LDOS3D}
\end{figure*}

\begin{figure*}
    \centering
    {\hspace*{-2em}\includegraphics[width = 8.6cm]{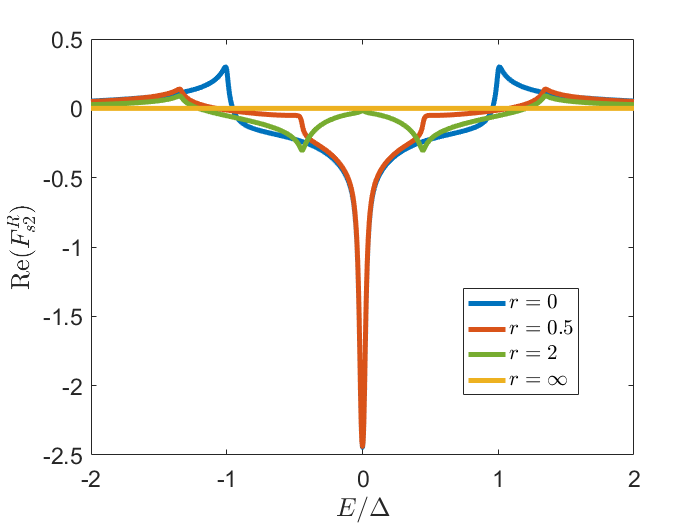}}
    \hfill
    {\hspace*{-2em}\includegraphics[width = 8.6cm]{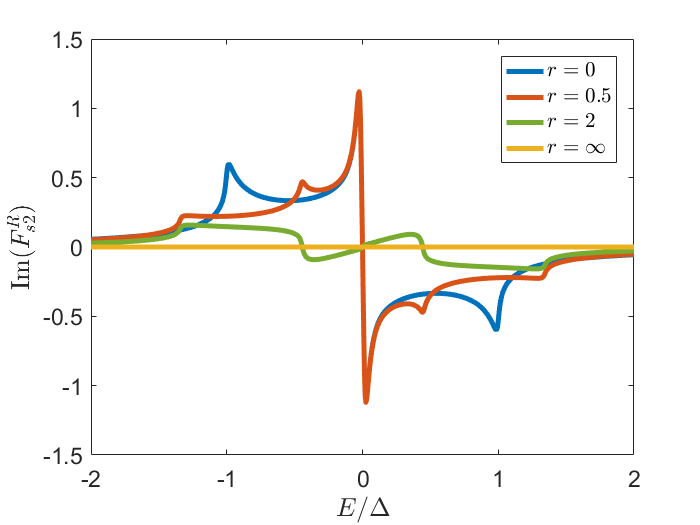}}
    \hfill
    {\hspace*{-1.5em}\includegraphics[width = 8.6cm]{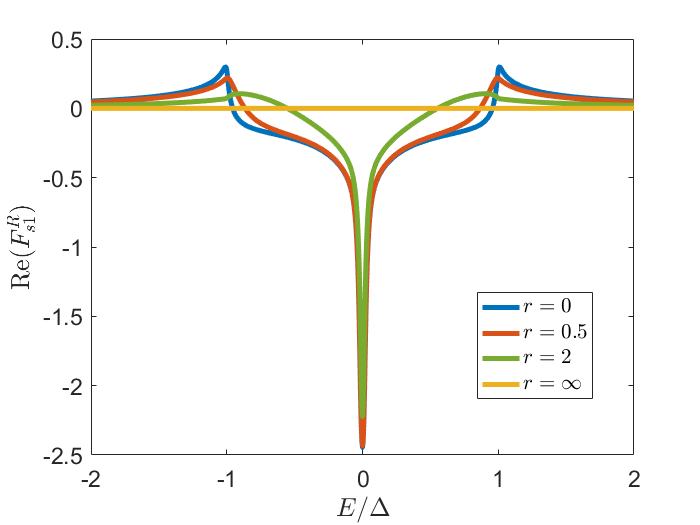}}
    \hfill
    {\hspace*{-2em}\includegraphics[width = 8.6cm]{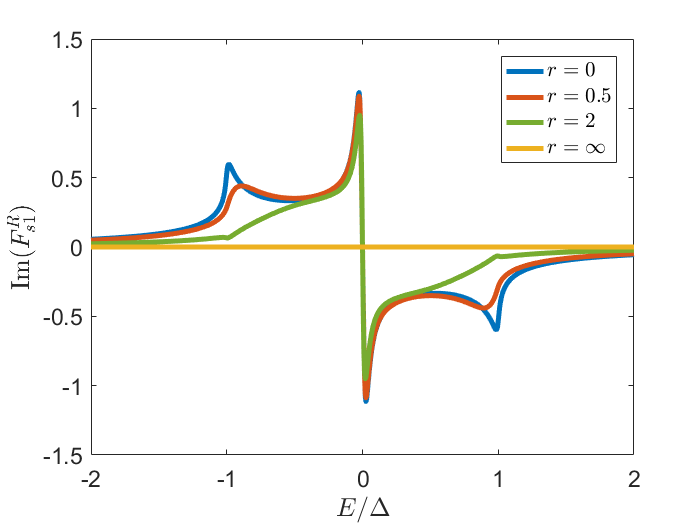}}
    \caption{The singlet pair amplitude for s+p-wave (top) and is+p-wave (bottom) superconductors for s-wave dominant  and p-wave dominant  s+B-W superconductors.}\label{fig:PAS3D}
\end{figure*}

\begin{figure*}
    \centering
    {\hspace*{-2em}\includegraphics[width = 8.6cm]{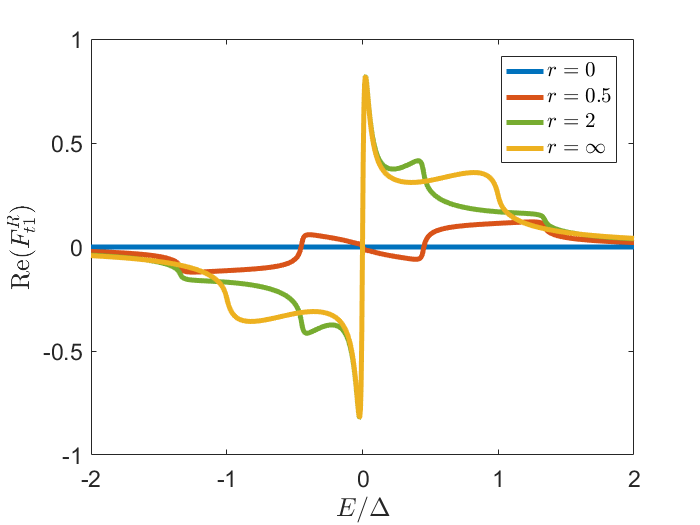}}
    \hfill
    {\hspace*{-2em}\includegraphics[width = 8.6cm]{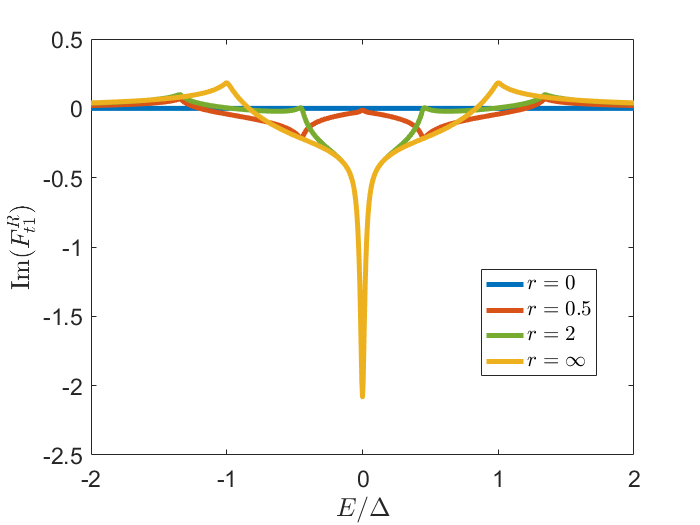}}
    \hfill
    {\hspace*{-1.5em}\includegraphics[width = 8.6cm]{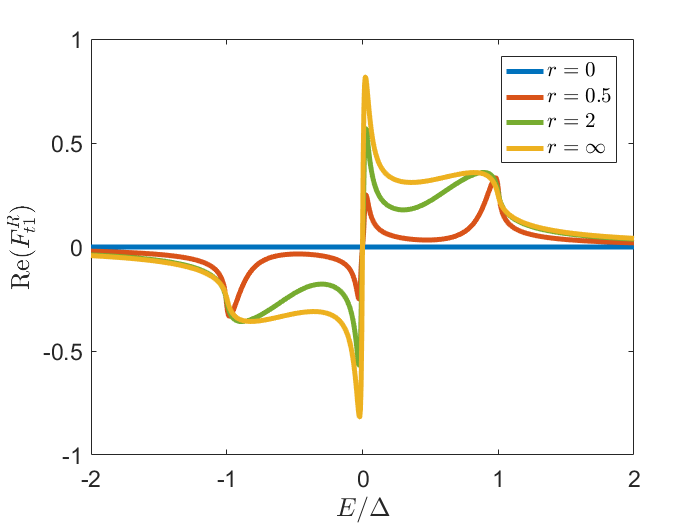}}
    \hfill
    {\hspace*{-2em}\includegraphics[width = 8.6cm]{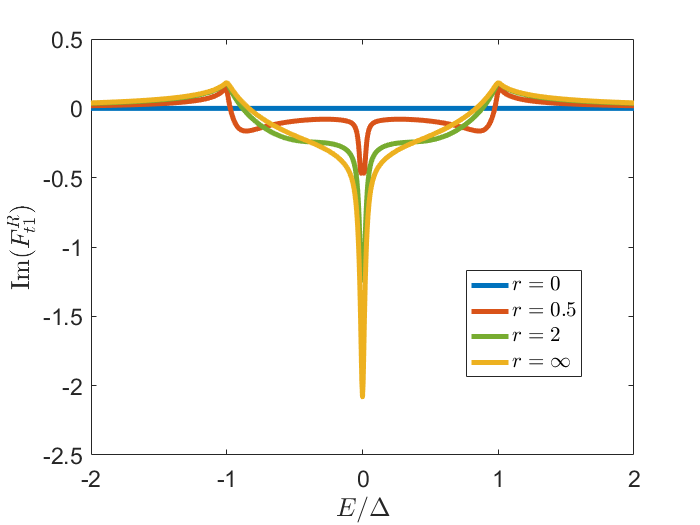}}
    \caption{The triplet pair amplitude for s+p-wave (top) and is+p-wave (bottom) superconductors for s-wave dominant  and (B-W) p-wave dominant  superconductors.}\label{fig:PAT3D}
\end{figure*}

In this section, we present the density of states and pair amplitudes in junctions with (i)s+B-W superconductors. We find that the results are similar to those for the two-dimensional junctions with s+helical p-wave superconductors.
The density of states and pair amplitudes are shown in Figs. \ref{fig:LDOS3D}, \ref{fig:PAS3D} and \ref{fig:PAT3D} respectively. The results are similar to the 2D helical superconductor. For the s+p-wave superconductor, the zero energy density of states is independent of the subdominant component, it depends only on whether the s-wave or p-wave component of the pair potential is dominant. On the other hand, for the is+p-wave superconductor, the zero energy density of states continuously changes from the s-wave to the p-wave value, being lower than the normal density of states for $r = 2$.

Because we consider only s+p and is+p wave superconductors the phases of the singlet and triplet components of the pair potential are independent of energy and either only the $\tau_{1}$-component or only the $\tau_{2}$-component is nonzero. In the following we show only those pair amplitudes which are not identically zero for all $r$.
The singlet pair amplitude is absent for p-wave superconductors, and has a large peak at $E = 0$ with two smaller peaks at $E = \pm\Delta_{0}$ for s-wave superconductors. If the s-wave component of the pair potential is dominant the induced singlet pair amplitude peaks at $E = 0$, having exactly (s+p-wave) or approximately (is+p-wave) the same value at zero energy as for the s-wave superconductor. 

For s+p-wave superconductors there are peaks at $|E| = \Delta_{s}\pm\Delta_{t}$, whereas for is+p-wave superconductors there are only broader peaks at $|E| = \Delta_{0}$. For p-wave dominant junctions the singlet pair amplitude strongly depends on the relative phase between the s-wave and p-wave components of the pair potential. A common feature it that the singlet pair amplitude is always smaller than in s-wave dominant junctions. However, for s+p-wave superconductors the singlet pair amplitude vanishes at $E = 0$, whereas for is+p-wave junctions the singlet pair amplitude has a peak at zero energy.
For the triplet pair amplitude the reverse holds. It vanishes for s-wave superconductors, peaks around $E = 0$ for p-wave or p-wave dominant superconductors, and for s-wave dominant superconductors the result highly depends on the phase difference between the s-wave and p-wave components of the pair potential. For s+p-wave superconductors it vanishes at $E = 0$, for is+p-wave superconductors the triplet pair amplitude has a peak at zero energy, though it is much smaller than for p-wave 
(dominant) superconductors.

\end{document}